\documentclass[bibyear]{aa}  
\usepackage[justification=centering]{caption}
\usepackage[rightcaption]{sidecap}
\usepackage{subcaption}
\usepackage{graphicx}
\usepackage{natbib}
\usepackage{color}
\usepackage{threeparttable}
\newcommand{\arcsecond}{^{\prime\prime}}
\usepackage[singlelinecheck=false,justification=justified]{caption}
\usepackage{txfonts}

\usepackage{gensymb}

\usepackage{hyperref}
\hypersetup{
    colorlinks=true, % Enable colored links
    linkcolor=blue, % Internal document links color
    citecolor=blue, % Citation links color
    urlcolor=blue % URL links color
}
\usepackage{mathtools}
\usepackage{bm}
\usepackage{esvect}
\begin{document}

    \title{Stellar obliquities of eight close-in gas giant exoplanets}
    %\subtitle{Spin-orbit measurements of eight gas giants}
    \titlerunning{Stellar obliquities of eight close-in gas giant exoplanets}
    
   \author{J. Zak
          \inst{1, 2, 3}
          \and
          H.\,M.\,J. Boffin\inst{3}
           \and
          E. Sedaghati\inst{4}
           \and
          A. Bocchieri\inst{5}
            \and
          Z. Balkoova\inst{1}
            \and
          M. Skarka\inst{1, 6}
            \and
           P. Kabath \inst{1}}

\institute{
    Astronomical Institute of the Czech Academy of Sciences, Fri\v{c}ova 298, 25165 Ond\v{r}ejov, Czech Republic;
    \email{zak@asu.cas.cz}
    \and
    Faculty of Physics and Astronomy, Friedrich-Schiller-Universität, Fürstengraben 1, 07743, Jena, Germany
    \and
    European Southern Observatory, Karl-Schwarzschild-str. 2, 85748 Garching, Germany
    \and
    European Southern Observatory, Casilla 13, Vitacura, Santiago, Chile
    \and
    Dipartimento di Fisica, La Sapienza Università di Roma, Piazzale Aldo Moro 5, Roma, 00185, Italy
    \and
    Department of Theoretical Physics and Astrophysics, Masaryk Univesity, Kotl\'a\v{r}sk\'a 2, 60200 Brno, Czech Republic\\
}

   \date{Received September 8, 2024; accepted October XX, 2024}

% \abstract{}{}{}{}{} 
% 5 {} token are mandatory

  \abstract
   {The Rossiter-McLaughlin effect allows us to measure the projected stellar obliquity of exoplanets. From the spin-orbit alignment, planet formation and migration theories can be tested to improve our understanding of the currently observed exoplanetary population. Despite having the spin-orbit measurements for more than 200 planets, the stellar obliquity distribution is still not fully understood, warranting additional measurements to sample the full parameter space. We analyze archival HARPS and HARPS-N spectroscopic transit time series of eight gas giant exoplanets on short orbits and derive their projected stellar obliquity $\lambda$. We report a prograde, but misaligned orbit for HAT-P-50b ($\lambda =41^\circ\ ^{+10}_{-9}$), possibly hinting at previous high-eccentricity migration given the presence of a close stellar companion. We measured sky-projected obliquities that are consistent with aligned orbits for the rest of the planets: WASP-48b ($\lambda =-4^\circ\ \pm$ 4), WASP-59b ($\lambda =-1^\circ\ ^{+20}_{-21}$), WASP-140 Ab ($\lambda =-1^\circ\ \pm$ 3), WASP-173 Ab ($\lambda =9^\circ\ \pm$ 5), TOI-2046b ($\lambda =1^\circ\ \pm$ 6), HAT-P-41 Ab ($\lambda =-4^\circ\ ^{+5}_{-6}$), and Qatar-4b ($\lambda =-13^\circ\ ^{+15}_{-19}$).    
   We measure the true stellar obliquity $\psi$ for four systems. We infer a prograde, but misaligned, orbit for TOI-2046b  with $\psi=$42$^{+10}_{-8}$\,deg.  Additionally, $\psi = 30^\circ\ ^{+18}_{-15}$ for WASP-140 Ab, $\psi = 21^\circ\ ^{+9}_{-10}$ for WASP-173 Ab,  and $\psi = 32^\circ\ ^{+14}_{-13}$ for Qatar-4b. The aligned orbits are consistent with slow disk migration, ruling out violent events that would excite the orbits over the history of these systems. Finally, we provide a new age estimate for TOI-2046 of at least 700 Myr and for Qatar-4 of at least 350-500 Myr, contradicting previous results.}


   \maketitle
%
%________________________________________________________________

\section{Stellar obliquity of exoplanets} \label{sec:intro}

Linking the observed exoplanetary population to the conditions in the protoplanetary disk is at the frontier of exoplanetary science. Exoplanetary atmospheres can provide a myriad of information on the system's history \citep{madhu19,tur21}. A complementary approach is to study the formation and evolution of exoplanetary systems through their dynamical parameters, i.e., eccentricity, orbital period, and stellar obliquity\footnote{The measured orbital inclination is not an intrinsic property of the system as it depends on the line of sight.} \citep[e.g.,][]{tur20,gaj23}: we can indeed attempt to link the observed parameters of the planetary systems to the properties of their host stars and the environments in which they have formed \citep[e.g.,][]{ob11,best22,eis23,gup24}. In order to perform these studies with high significance, we need to measure these quantities over a wide range of parameter space. 

The projected stellar obliquity ($\lambda$) is most commonly measured through the Rossiter-McLaughlin (R-M) effect -- a spectroscopic anomaly that occurs when the disk of a companion passes in front of the spinning star. Currently, there are more than 215 measurements\footnote{Retrieved from the TEPCat \citep{south11} on July 31st, 2024.} of the projected obliquity. Initially, the obliquity had been measured for hot Jupiters only, but advances in instrumentation \citep{ma03,pepe21,sei22} have allowed the measurement of the spin-orbit alignment for Neptunian, sub-Neptunian and even terrestrial planets.
Such studies have revealed a complex architecture of exoplanetary systems with planetary orbits varying from well-aligned to slightly misaligned, and to planets on polar and even retrograde orbits. A more detailed description of the R-M effect is given in \citet{tri17} and \citet{alb22}.

Misaligned and polar orbits are often thought of as indications of violent history within the system as they contradict the standard formation processes where the planet and the star form from the same rotating disk. Hence, to explain these orbits, violent mechanisms, such as long-term interactions with a binary companion \citep{bat12}, close fly-bys of other stars \citep{bat10},  shifts of the stellar rotation axis relative to the disk due to the stellar magnetic field \citep{lai11}, gravitational scatterings among the planets \citep{chat08}, or long-term perturbations due to a companion caused by Kozai-Lidov mechanism \citep{fab07}, are required.

The majority of stars are formed in binary or higher-multiplicity systems \citep{rag10,elb24}.
In addition to the spin-orbit orientation (between a planet and its host star\footnote{For clarity, in this work we talk only about s-type planets, that are those that orbit around one star in a binary pair.}), orbit-orbit orientation (between the two components in a binary system) allows for more in-depth investigation of how the binary system could have formed \citep[e.g.,][]{rice22}. Several trends were identified in the exoplanetary population around host stars in binary pairs. In particular, \citet{su21} identified a population of massive short-period planets in close binaries that are almost absent in wide binaries or single stars. Some of the proposed explanations of this trend include enhanced planetary migration, collisions, and/or ejections in close binaries. Those are the same mechanisms that are also suggested for exciting the stellar obliquities of planetary systems. However, the exact role of stellar companions and planetary disks is not fully understood and requires further investigation to apprehend their interplay and the observed population \citep{bar24}.

The unprecedented astrometric sensitivity of the \textit{Gaia} mission \citep{gaia23} has allowed for the systematic study of both spin-orbit and orbit-orbit study \citep{chris24}. \citet{rice24} have studied the obliquity of 48 systems hosting an exoplanet and being part of a binary or triple system. From the observed line-of-sight orbit-orbit alignment and under-abundance of face-on systems that correspond to near-polar binary configuration, they suggest this might be due to viscous dissipation induced by nodal recession during the protoplanetary disk phase. Furthermore, they report that there is no clear correlation between spin-orbit misalignment and orbit-orbit misalignment, as initially reported by \citet{beh22}. However, further measurements are needed, as the sample is still quite limited.

In this study, we present, based on unpublished HARPS and HARPS-N archival data, the projected obliquity measurements of eight gas giants on short orbits around FGK dwarfs, with at least five of them being part of binary or triple systems and three astrometrically resolved.

\begin{table*}[htbp]
\captionsetup{justification=centering}
\caption{Observing logs for all eight planets. The number in parenthesis represents the number of frames taken in-transit.}
\vspace{-.4cm}
\label{obs_logs}
\centering
\begin{center}
\begin{tabular}{lcccccc}
\hline
Target & Night & No.     & Exp.     & Airmass & Median & Instrument \\ 
     & Obs~&~frames~& Time (s) & range    &  S/N$^a$    \\ 
\hline
WASP-48 & 2018-06-20/21~&~24 (17)~& 900     & 1.23-1.12-1.20~&~23-33 & HARPS-N \\ 

WASP-59 & 2016-09-04/05~&~29 (14)~& 600     & 1.06-1.00-1.42~&~9-15& HARPS-N \\ 

WASP-140 A & 2018-10-02/03~&~24 (9)~& 600     &1.35-1.01-1.06~&~14-31& HARPS\\
 & 2018-11-27/28~&~20 (7)~& 900     &1.15-1.01-1.37~&~39-51& HARPS\\

WASP-173 A & 2018-10-18/19~&~35 (13)~& 600     & 1.10-1.00-1.59~&~17-27 & HARPS \\

TOI-2046 & 2022-09-29/30~&~31 (13)~& 600     & 2.07-1.43~&~15-26 & HARPS-N \\

HAT-P-41 A & 2016-07-05/06~&~34 (18)~& 800     & 2.17-1.09-1.75~&~29-48& HARPS-N\\  
 & 2016-07-13/14~&~33 (17)~& 900     & 2.24-1.09-2.17~&~20-42& HARPS-N\\ 

HAT-P-50 & 2016-12-14/15~&~37 (21)~& 900     & 1.45-1.44-1.70~&~12-16& HARPS-N \\

Qatar-4 & 2017-10-31/01~&~ 17 (8)~& 900 & 1.19-1.04-1.13~&~7-11& HARPS-N \\  

\hline 
\end{tabular}

\vspace{0.2cm}
\textit{} $^a$ S/N in the extracted 1-D spectrum, per spectral resolution element in the order at 550nm.
\end{center}
\end{table*}

\begin{table*}[htbp]
\caption{Properties of the targets (star and planet).}
\vspace{-.4cm}
\label{tab:targets1}
\begin{center}
\begin{tabular}{@{ }l@{ }l@{ }c@{ }c@{ }c@{ }c@{ }} 
\hline
& Parameters & WASP-48  & WASP-59  & WASP-140 A & WASP-173 A \\ 
\hline
\textbf{Star} & V mag & 11.66 & 13.00 & 11.1 & 11.3 \\
 & Sp. Type  & G0 & K5 & K0 & G3 \\
 & M$_{\rm{s}}$ (M$_\odot$) & 1.09 $\pm$ 0.08 & 0.719 $\pm$ 0.035 & 0.90 $\pm$ 0.04 & 1.05 $\pm$ 0.08 \\
 & R$_{\rm{s}}$ (R$_\odot$) & 1.09 $\pm$ 0.14 & 0.613 $\pm$ 0.044 & 0.87 $\pm$ 0.04 & 1.11 $\pm$ 0.05 \\
 & T$_{\rm{eff}}$ (K) & 6000 $\pm$ 150 & 4650 $\pm$ 150 & 5300 $\pm$ 100 & 5700 $\pm$ 150 \\
 & $v\,\rm{sin}i_*$ (km/s) & 12.2 $\pm$ 0.7 & 2.3 $\pm$ 1.2 & 3.1 $\pm$ 0.8 & 6.1 $\pm$ 0.3 \\
\textbf{Planet} & M$_{\rm{p}}$ (M$_{\rm Jup}$) & 0.98 $\pm$ 0.09 & 0.863 $\pm$ 0.045 & 2.44 $\pm$ 0.07 & 3.69 $\pm$ 0.18 \\
 & R$_{\rm{p}}$ (R$_{\rm Jup}$) & 1.67 $\pm$ 0.08 & 0.775 $\pm$ 0.068 & 1.44$^{+0.42}_{-0.18}$ & 1.20 $\pm$ 0.06 \\
 & Period (d) & 2.143634 $\pm$ 0.000003 & 7.919585 $\pm$ 0.000010 & 2.2359835 $\pm$ 0.0000008 & 1.38665318 $\pm$ 0.00000027 \\
 & $\rm{T_0}$ - 2450000 (d) & 5364.55043 $\pm$ 0.00027 & 5830.95559 $\pm$ 0.00053 & 6912.35105 $\pm$ 0.00015 & 7288.8585 $\pm$ 0.0002 \\
 & a (AU) & 0.03444 $\pm$ 0.00043 & 0.0697 $\pm$ 0.0011 & 0.0323 $\pm$ 0.0005 & 0.0248 $\pm$ 0.0006 \\
 & e & 0 & 0.100 $\pm$ 0.042 & 0.0470 $\pm$ 0.0035 & 0 \\
 & i (deg) & 80.09$^{+0.69}_{-0.55}$ & 89.27 $\pm$ 0.52 & 83.3$^{+0.5}_{-0.8}$ & 85.2 $\pm$ 1.1 \\
 & T$_{\rm{eq}}$ (K) & 2030 $\pm$ 70 & 670 $\pm$ 35 & 1320 $\pm$ 40 & 1880 $\pm$ 55 \\
 & Discovery ref. & \citet{eno11} & \citet{heb13} & \citet{hel17} & \citet{hel19} \\
  & Prior ref. & \citet{mac22} & Appendix \ref{sec:w59photo} & \citet{ale22} & \citet{hel19} \\
\hline
\end{tabular}

\vspace{0.4cm}

\begin{tabular}{@{ }l@{ }l@{ }c@{ }c@{ }c@{ }c@{ }} 
\hline
& Parameters & TOI-2046 & HAT-P-41 A & HAT-P-50 & Qatar-4 \\ 
\hline
\textbf{Star} & V mag & 11.55 & 11.09 & 11.76 & 13.6 \\
 & Sp. Type  & F8 & F6 & F7 & K1 \\
 & M$_{\rm{s}}$ (M$_\odot$) & 1.31 $\pm$ 0.09 & 1.418 $\pm$ 0.047 & 1.273 $\pm$ 0.115 & 0.896 $\pm$ 0.048 \\
 & R$_{\rm{s}}$ (R$_\odot$) & 1.21 $\pm$ 0.07 & 1.19 $\pm$ 0.08 & 1.698 $\pm$ 0.071 & 0.849 $\pm$ 0.063 \\
 & T$_{\rm{eff}}$ (K) & 6160 $\pm$ 100 & 6390 $\pm$ 100 & 6280 $\pm$ 49 & 5198 $\pm$ 42 \\
 & $v\,\rm{sin}i_*$ (km/s) & 9.8 $\pm$ 1.6 & 19.6 $\pm$ 0.5 & 8.90 $\pm$ 0.50 & 7.1 $\pm$ 0.3 \\
\textbf{Planet} & M$_{\rm{p}}$ (M$_{\rm Jup}$) & 1.13 $\pm$ 0.19 & 0.800 $\pm$ 0.102 & 1.350 $\pm$ 0.073 & 5.36 $\pm$ 0.20 \\
 & R$_{\rm{p}}$ (R$_{\rm Jup}$) & 1.21 $\pm$ 0.07 & 1.685$^{+0.076}_{-0.051}$ & 1.288 $\pm$ 0.064 & 1.135 $\pm$ 0.110 \\
 & Period (d) & 1.4971842 $\pm$ 0.0000031 & 2.694047 $\pm$ 0.000004 & 3.122019 $\pm$ 0.000063 & 1.8053564 $\pm$ 0.0000043 \\
 & $\rm{T_0}$ - 2450000 (d) & 7792.2767 $\pm$ 0.0024 & 4983.86167 $\pm$ 0.00107 & 56285.99093 $\pm$ 0.00034 & 7637.77370 $\pm$ 0.00046 \\
 & a (AU) & 0.0248 $\pm$ 0.0006 & 0.0426 $\pm$ 0.0005 & 0.04530 $\pm$ 0.00069 & 0.02803 $\pm$ 0.00048 \\
 & e & 0 & 0 & 0 & 0.057$^{+0.022}_{-0.018}$ \\
 & i (deg) & 83.6 $\pm$ 0.9 & 87.7 $\pm$ 1.0 & 83.65 $\pm$ 0.57 & 87.5 $\pm$ 1.6 \\
 & T$_{\rm{eq}}$ (K) & 1900 $\pm$ 70 & 1941 $\pm$ 38 & 1862 $\pm$ 34 & 1385 $\pm$ 50 \\
 & Discovery ref. & \citet{kab22} & \citet{hart12} & \citet{hart15} & \citet{als17} \\
  & Prior ref. & \citet{kab22} & \citet{hart12} & \citet{hart15} & \citet{als17} \\
\hline
\end{tabular}

\end{center}
\end{table*}

\section{Data sets and their analyses} \label{sec:data}
Our datasets come from two sources: two datasets (WASP-140~A and WASP-173 A) come from the HARPS instrument \citep{ma03} mounted at the ESO 3.6-m telescope at La Silla, Chile. The rest of the targets come from the HARPS-N instrument \citep{cos12}, mounted at the 3.58-m TNG telescope at La Palma, Canary Islands. The HARPS data were obtained from the ESO science archive\footnote{\url{http://archive.eso.org}}: program IDs 0102.C-0618(A), PI: Esposito; 0102.C-0319(A), PI: Anderson; while the HARPS-N data were obtained from the TNG archive\footnote{\url{http://archives.ia2.inaf.it/tng/}}: program IDs CAT16B-119, PI: González; CAT17B-155, PI: Murgas Alcaíno; CAT18A-S, PI: IAC Service; OPT16A-49, PI: Ehrenreich and CAT22A-9, PI: Orell. Table \ref{obs_logs} shows the properties of the data sets that were used. The downloaded data are fully reduced products as processed with the HARPS Data Reduction Software (DRS version 3.8). Each spectrum is provided as a merged 1D spectrum resampled onto a 0.001 nm uniform wavelength grid. The wavelength coverage of the spectra spans from 380 to 690 nm, with a resolving power of R\,$\approx$\,115\,000, corresponding to 2.7~km\,s$^{-1}$ per resolution element. The spectra are already corrected to the Solar system barycentric frame of reference. We further summarize the properties of the studied systems in Table \ref{tab:targets1}.

\subsection{Rossiter–McLaughlin effect}

\begin{figure}[h]
\includegraphics[width=0.45\textwidth]{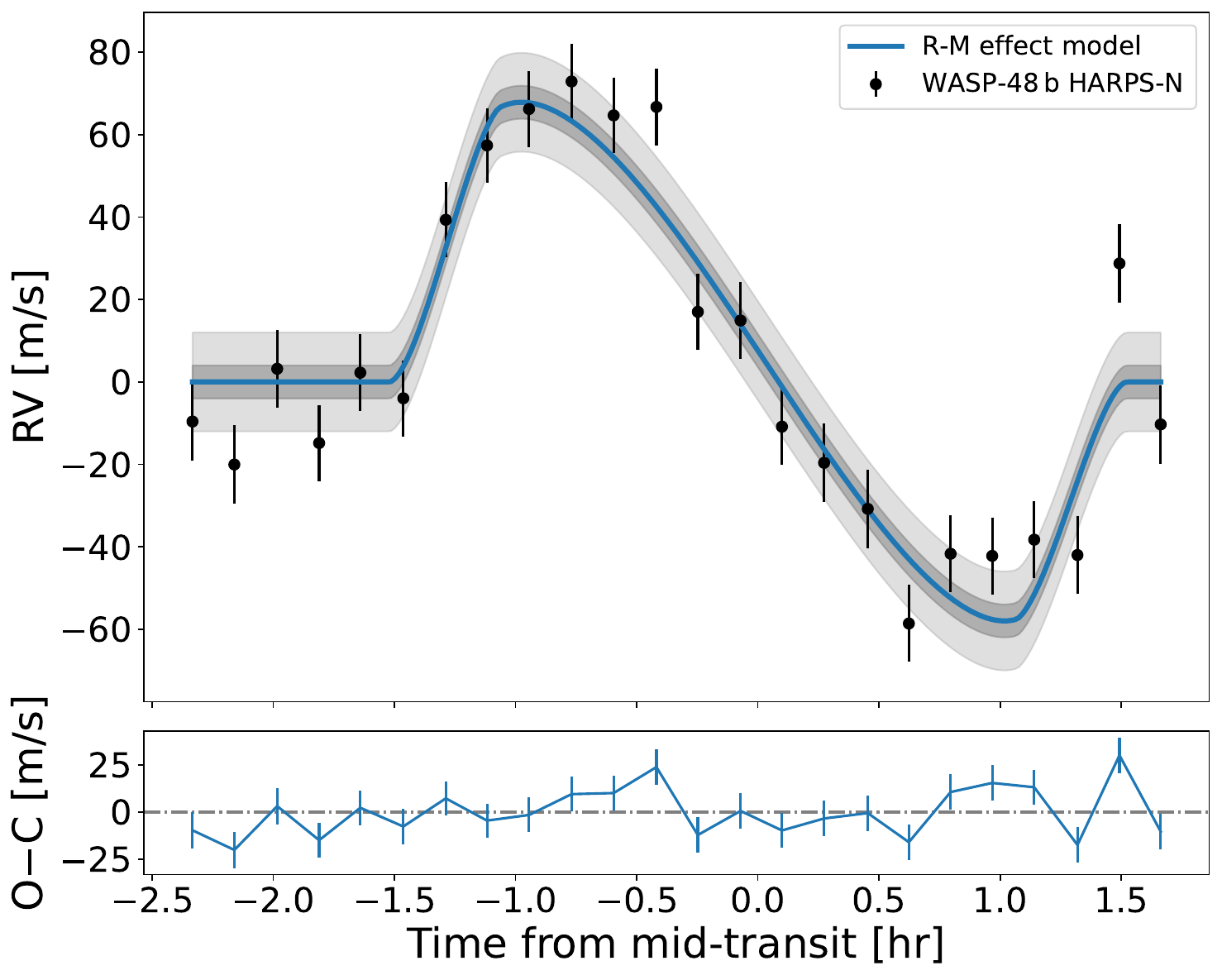}
\caption{The Rossiter-McLaughlin effect of WASP-48b observed with HARPS-N. The observed data points (black) are shown with their error bars. The systemic and Keplerian orbit velocities were removed. The blue line shows the best fitting model to the data, together with 1-$\sigma$ (dark grey) and 3-$\sigma$ (light grey) confidence intervals.}
\label{f:48}
\end{figure}

\begin{figure}
\includegraphics[width=0.45\textwidth]{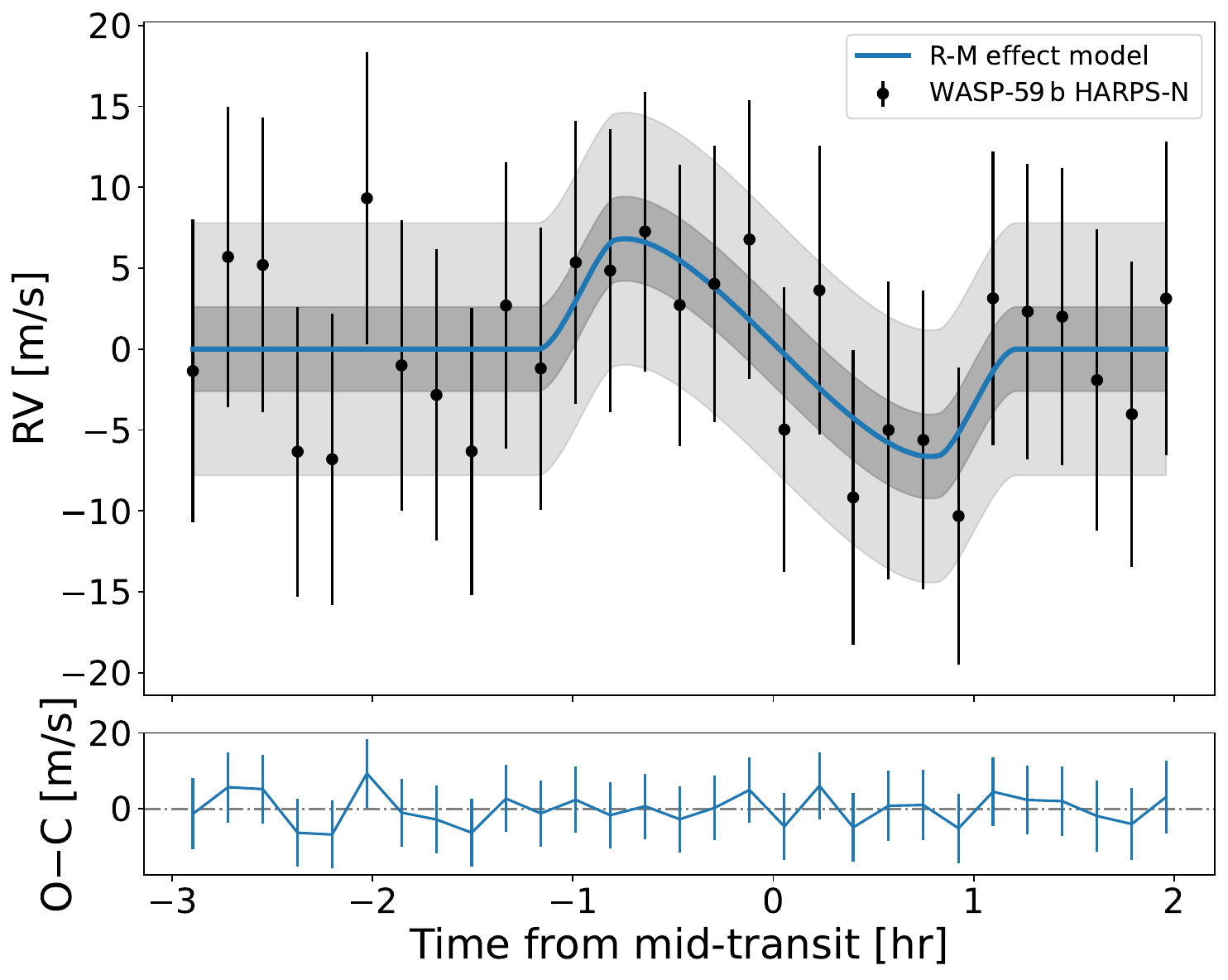}
\caption{Same as Fig.~\ref{f:48} for WASP-59b.}
\label{f:59}
\end{figure}

\begin{figure}
\includegraphics[width=0.45\textwidth]{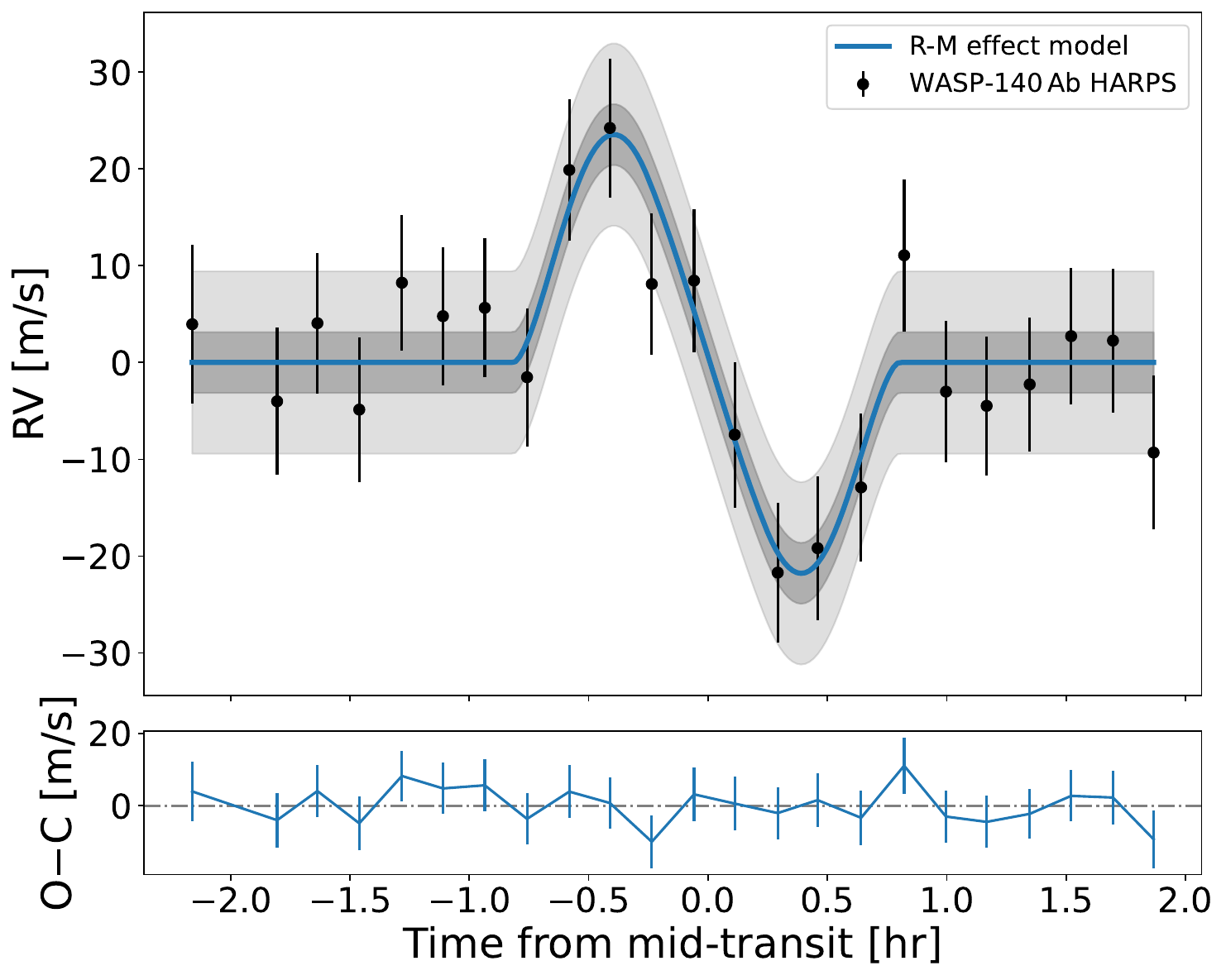}
\caption{Same as Fig.~\ref{f:48} for WASP-140 Ab with HARPS.}
\label{f:140}
\end{figure}

\begin{figure}
\includegraphics[width=0.45\textwidth]{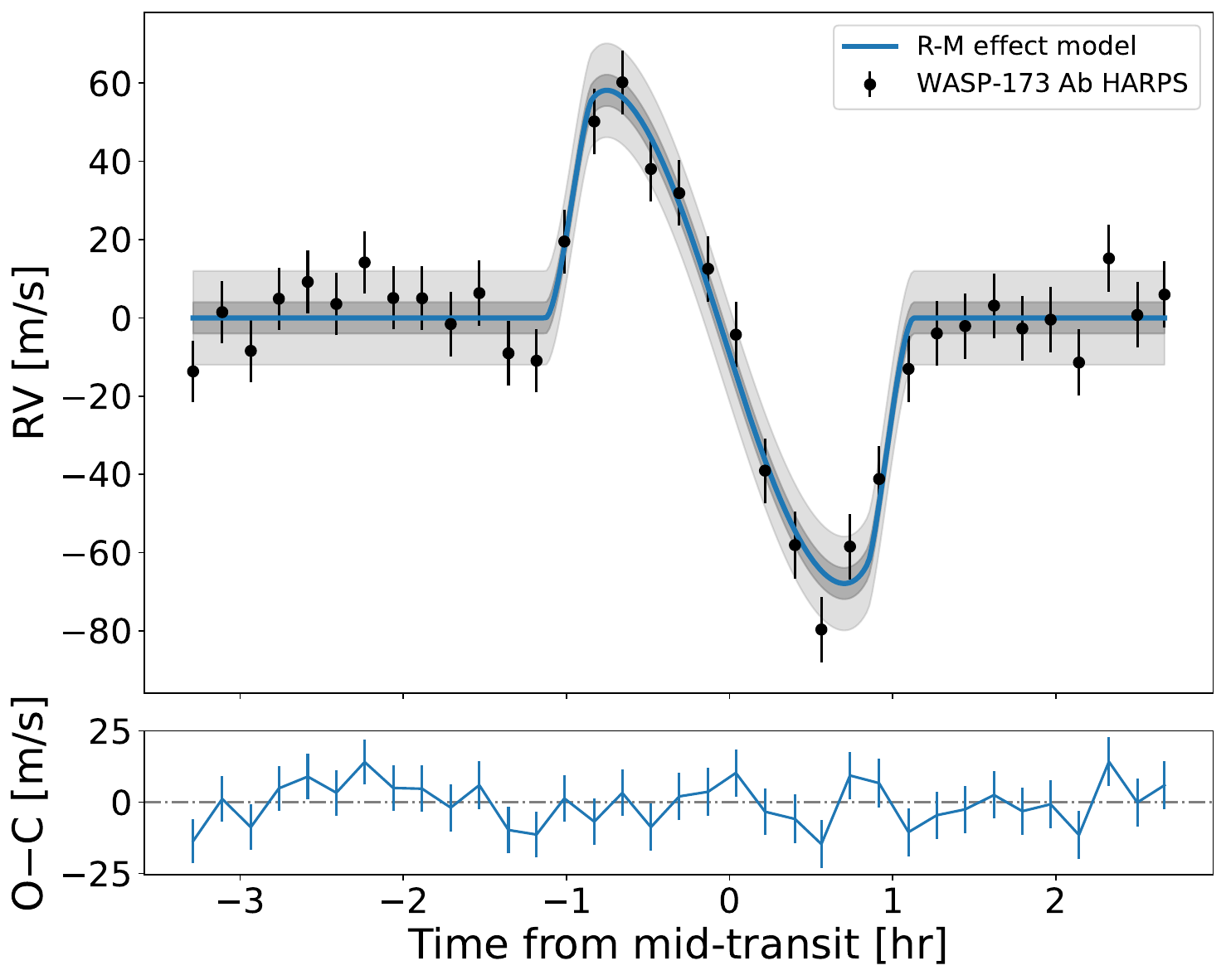}
\caption{Same as Fig.~\ref{f:48} for WASP-173 Ab with HARPS.}
\label{f:173}
\end{figure}

\begin{figure}
\includegraphics[width=0.45\textwidth]{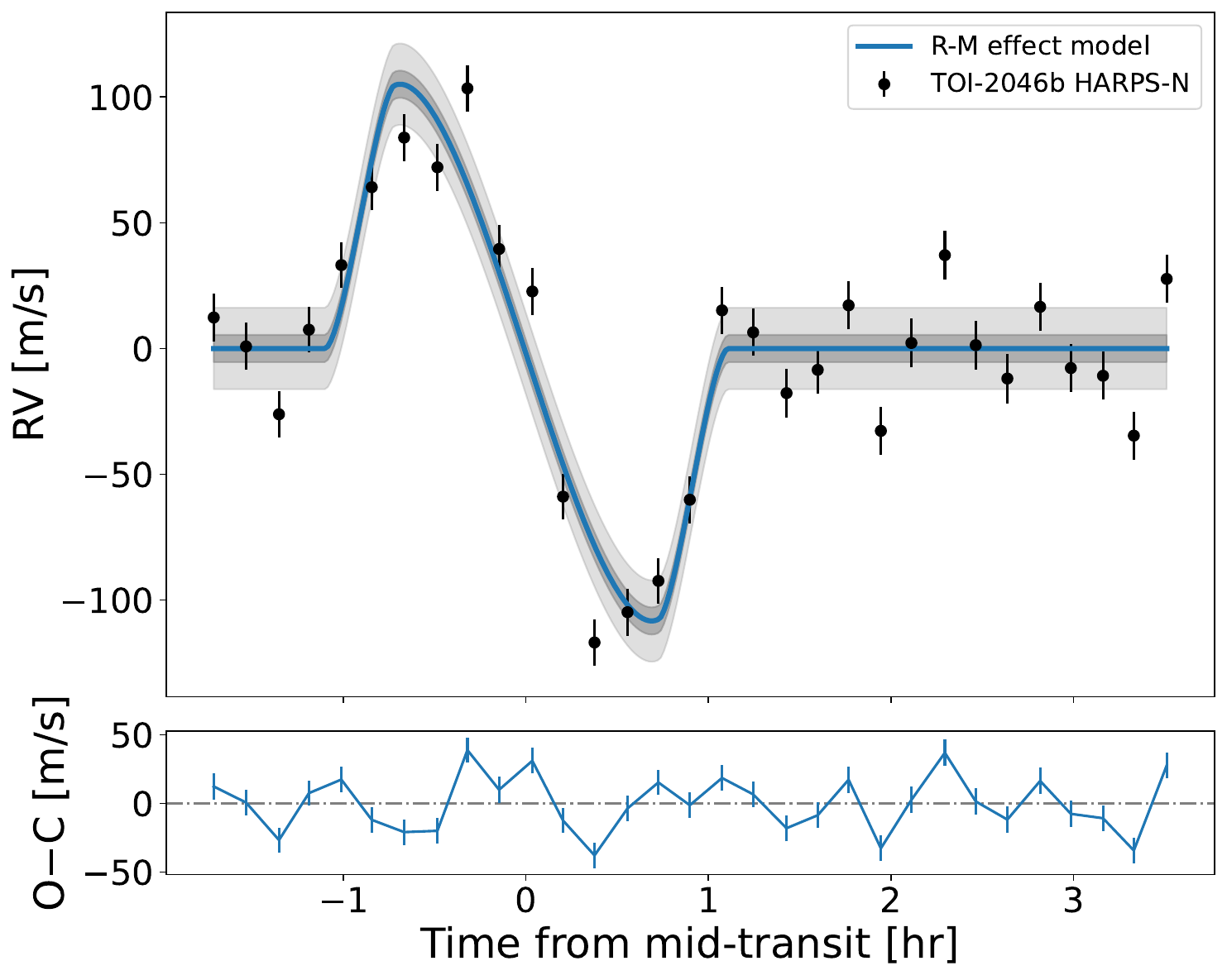}
\caption{Same as Fig.~\ref{f:48} for TOI-2046b.}
\label{f:t2046}
\end{figure}

\begin{figure}
\includegraphics[width=0.45\textwidth]{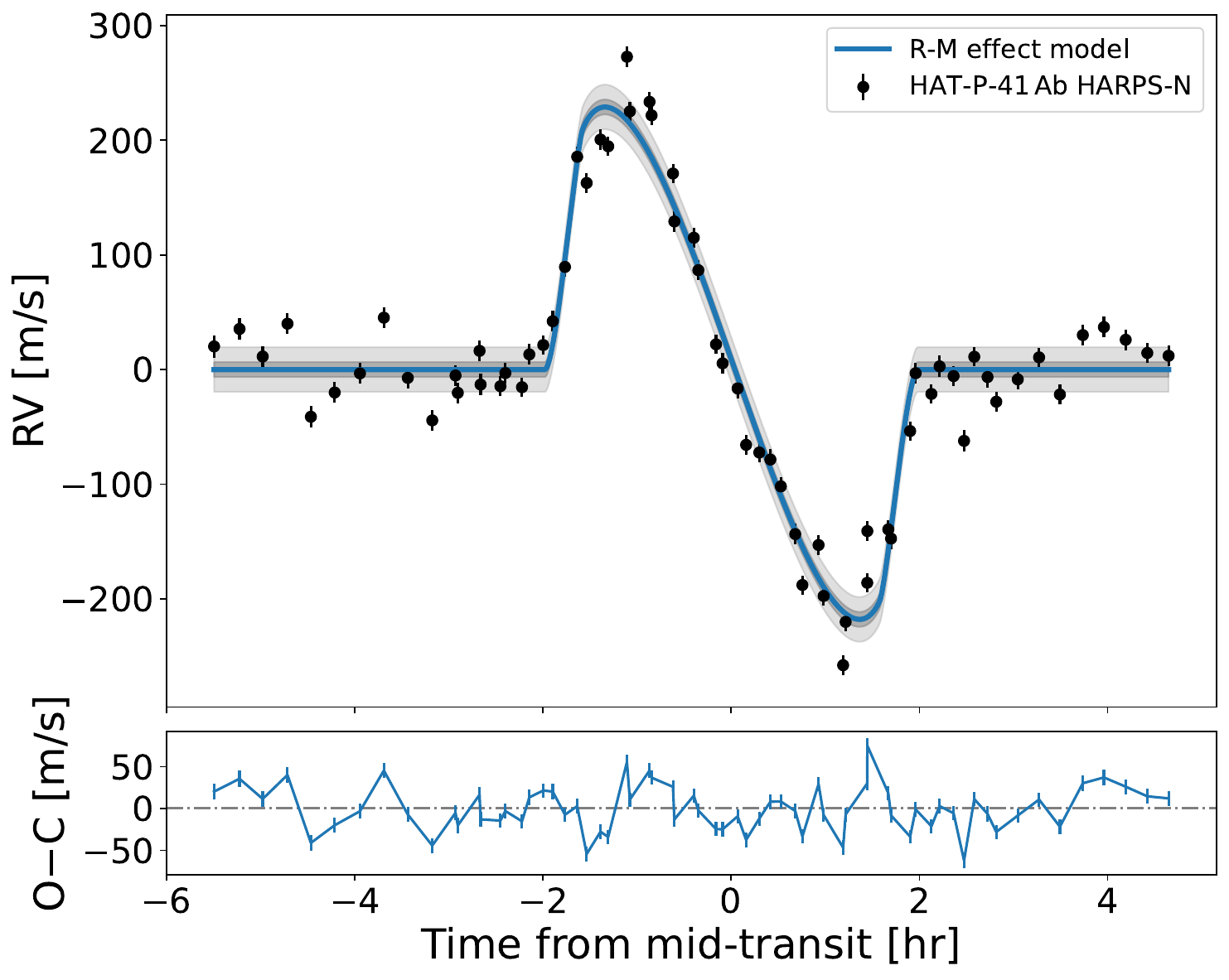}
\caption{Same as Fig.~\ref{f:48} for HAT-P-41 Ab.}
\label{f:hp41}
\end{figure}

\begin{figure}
\includegraphics[width=0.45\textwidth]{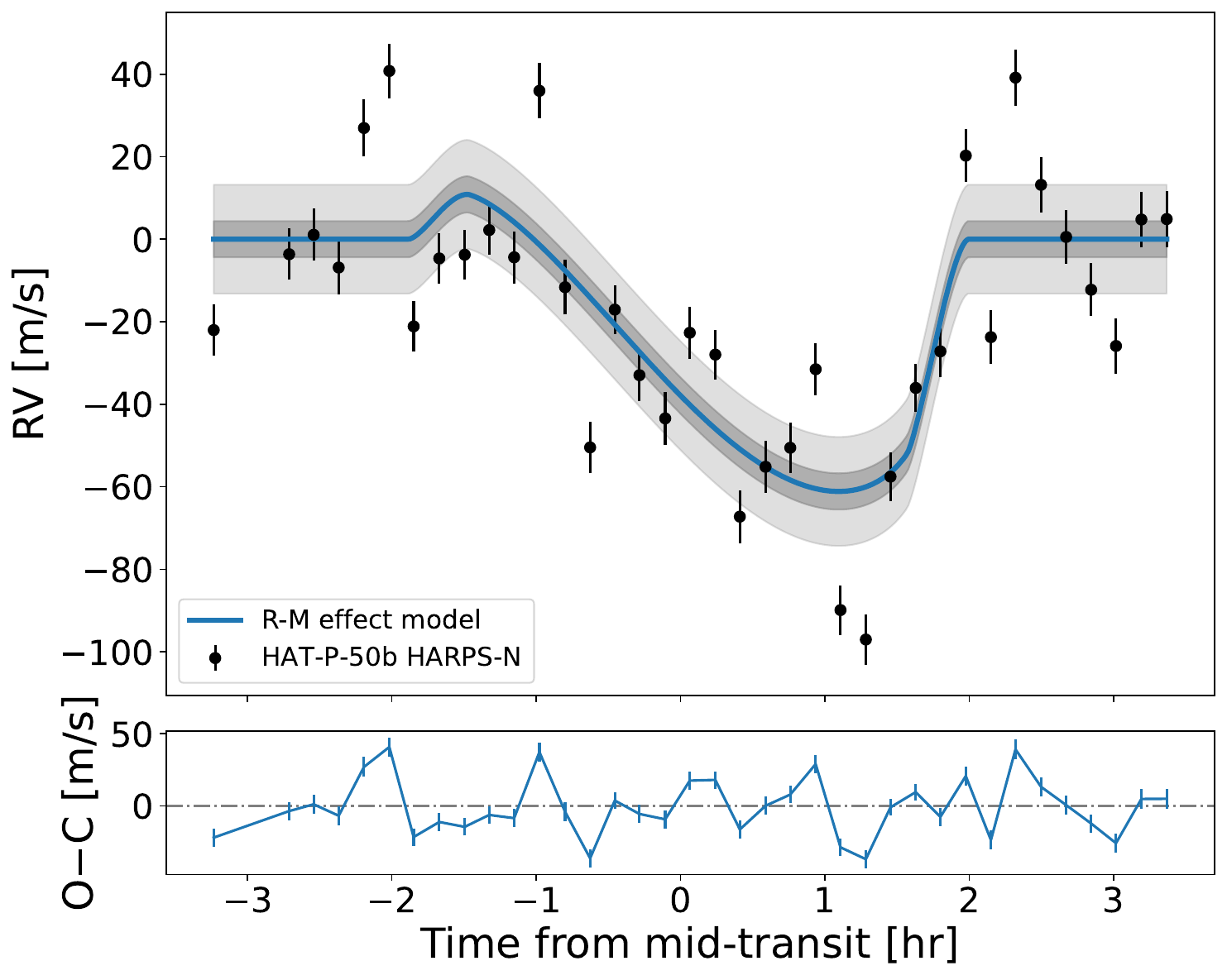}
\caption{Same as Fig.~\ref{f:48} for HAT-P-50b.}
\label{f:hp50}
\end{figure}

\begin{figure}
\includegraphics[width=0.45\textwidth]{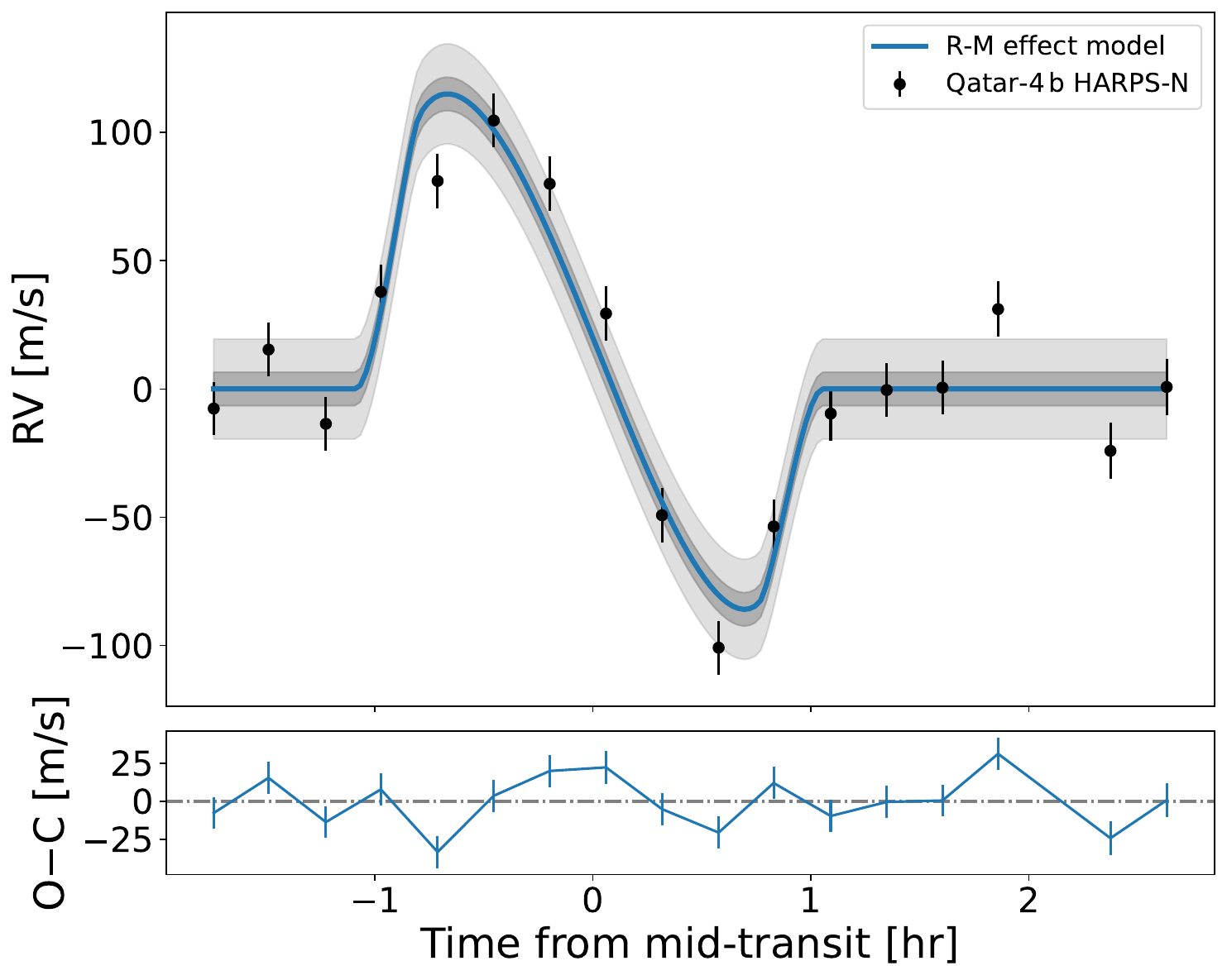}
\caption{Same as Fig.~\ref{f:48} for Qatar-4b.}
\label{f:q4}
\end{figure}

The R-M effect causes a spectral line asymmetry, or, equivalently, an asymmetry in the cross-correlation function (CCF) that manifests as anomalous radial velocities during the transit of the exoplanet. We have obtained the radial velocity measurements that are acquired from the HARPS and HARPS-N DRS pipelines together with their uncertainties. We have combined the results from two nights when available. To measure the projected stellar obliquity of the system ($\lambda$), we follow the methodology of \citet{zak24a}: we fit the RVs with a composite model, which includes a Keplerian orbital component as well as the R-M anomaly. This model is implemented in the \textsc{ARoMEpy}\footnote{\url{https://github.com/esedagha/ARoMEpy}} \citep{seda23} package, which utilizes the \textsc{Radvel} python module \citep{ful18} for the formulation of the Keplerian orbit. \textsc{ARoMEpy} is a Python implementation of the R-M anomaly described in the \textsc{ARoME} code \citep{bou13}. We used the R-M effect function defined for RVs determined through the cross-correlation technique in our code. We set Gaussian priors for the RV semi-amplitude ($K$) and the systemic velocity ($\Gamma$). In the R-M effect model, we fixed the following parameters to values reported in the literature: the orbital period ($P$), the planet-to-star radius ratio ($R_{\rm{p}}/R_{\rm{s}}$), and the eccentricity ($e$). The parameter $\sigma$, which is the width of the CCF and represents the effects of the instrumental and turbulent broadening, was measured on the data and fixed. Furthermore, we used the ExoCTK\footnote{\url{https://github.com/ExoCTK/exoctk}} tool to compute the quadratic limb-darkening coefficients with ATLAS9 model atmospheres \citep{cas03} in the wavelength range of the HARPS and HARPS-N instruments (380-690\,nm). We set Gaussian priors using the literature value and uncertainty derived from transit modeling (Tab.~\ref{tab:targets1}) on the following parameters during the fitting procedure\footnote{Except for WASP-59 for which we have analyzed previously unpublished \textit{TESS} photometry and present results in 
Appendix~\ref{sec:w59photo}.}: the central transit time ($T_C$), the orbital inclination ($i$), and the scaled semi-major axis ($a/R_{\rm{s}}$). Uniform priors were set on the projected stellar rotational velocity ($\nu\,\sin i_*$) and the sky-projected angle between the stellar rotation axis and the normal of the orbital plane ($\lambda$). To obtain the best fitting values of the parameters, we employed three independent Markov chain Monte Carlo (MCMC) ensemble simulations using the \textsc{Infer}\footnote{\url{https://github.com/nealegibson/Infer}} implementation using the Affine-Invariant Ensemble Sampler. We have initialized the MCMC at parameter values found by the Nelder-Mead method perturbed by a small value. We have used 20 walkers each with 12\,500 steps, burning the first 2\,500. As a convergence check, we ensured that the Gelman-Rubin statistic \citep{gel92} is less than 1.001 for each parameter. Using this setup we obtain our results and present them in 
Sect.~\ref{sec:resrm} and in Figs.~\ref{f:48} to \ref{f:q4}.

\section{Results} \label{sec:res}
\subsection{Projected stellar obliquity measurement}
\label{sec:resrm}

We have measured the projected stellar obliquity ($\lambda$) of seven targets for the first time: WASP-48b, WASP-59b, WASP-140 Ab, WASP-173 Ab, TOI-2046b, HAT-P-50b and Qatar-4b. HAT-P-41 Ab has been previously studied by \citet{john17}.

\textbf{WASP-48b} is an inflated hot Jupiter orbiting a G0V host star on a short 2.1-day orbit. The system potentially includes a wide stellar companion with $\Delta K_{\rm{s}} =7.3 \pm 0.1$ \citep{ngo16}. The planet's atmosphere has been studied by \citet{mur17}, who derived a flat transmission spectrum using the OSIRIS instrument mounted at the GTC. Recently, \citet{ben23} used the HPF spectrograph at the Hobby-Eberly Telescope to report a non-detection of the helium feature that hints at a low planetary mass-loss rate. In our analysis, we infer an aligned orbit of WASP-48b with $\lambda =-4 \pm 4$\,deg.

\textbf{WASP-59b} is a warm Jupiter orbiting a K5 host star on a 7.95-day orbit. \citet{fon21} identified, using \textit{Gaia} data, a wide ($>$ 9000 au) co-moving companion. We infer a prograde orbit with a broad but symmetric posterior in $\lambda$ around alignment, with $\lambda = -1_{-21}^{+20}$~deg.

\textbf{WASP-140 Ab} is a hot Jupiter orbiting a K0 host star on a 2.2-day orbit. The host star has a two magnitudes fainter companion, WASP-140 B, with a separation of \(7\arcsecond\).24 $\pm$ \(0\arcsecond\).01 and a position angle of 77.4 $\pm$ 0.1$^\circ$, corresponding to a projected separation of 850 au. We infer an aligned orbit of WASP-140 Ab with $\lambda =-1 \pm 3$\,deg.

\textbf{WASP-173 Ab} is a massive hot Jupiter orbiting a G3 host star on a 1.4-day orbit \citep{hel19,laba19}. The host star has a slightly less massive companion, WASP-173 B, with a separation of \(6\arcsecond\).1 $\pm$ \(0\arcsecond\).01 and a position angle of 110.1 $\pm$ 0.5$^\circ$, corresponding to a projected separation of 1440 au. We infer an aligned orbit of WASP-173 Ab with $\lambda =9 \pm 5$\,deg.

\textbf{TOI-2046b} is a hot Jupiter orbiting an F8 host star on a 1.5-day orbit. The reported age of the system between 100 and 400 Myrs as determined from gyrochronology and the lithium method \citep{kab22} would make it an amenable target for atmospheric escape study. Recently, \citet{orell24} reported upper limits for H$\alpha$ and He I excess absorption. We infer an aligned orbit of TOI-2046b with $\lambda =1 \pm 6$\,deg., and show below that the star is likely much older.

\textbf{HAT-P-41 Ab} is an inflated hot Jupiter. It orbits its F6 host star on a 2.7-day orbit. \citet{bohn20} confirmed the binary companion HAT-P-41 B with a separation of \(3\arcsecond\).621 $\pm$ \(0\arcsecond\).004 and a position angle of 183.9 $\pm$ 0.1$^\circ$, corresponding to a projected separation of 1240 au. \citet{shep21} reported high metal enrichment in the planetary atmosphere from  Hubble Space Telescope's (HST) transit spectroscopy; a result later confirmed by \citet{jia24} with GTC/OSIRIS. In our analysis, we infer an aligned orbit with $\lambda =-4.4^{+5.0}_{-5.6}$\,deg. The stellar obliquity of HAT-P-41 Ab has been previously studied with the Doppler tomography method by \citet{john17} using the High-Resolution Spectrograph mounted at the Hobby-Eberly Telescope. They have reported a prograde but slightly misaligned orbit with $\lambda =-22.1^{+0.8}_{-6.0}$\,deg. Our result points at a more aligned orbit, but both results are consistent within 3 sigmas. Furthermore, as noted in \citet{john17}, after the subtraction of their best fit model significant residuals remain and the authors have inflated their uncertainties by a factor of 2.5 to better account for possible systematics. We have applied the Doppler shadow method on HARPS-N data in Appendix \ref{DS} deriving $\lambda = -7.6^{+7.7}_{-8.0}$\,deg consistent with our classical R-M effect analysis.

\textbf{HAT-P-50b} is a hot Jupiter orbiting its F7 host star on a 3.1-day orbit \citep{hart15}.
\citet{les21} reported the presence of a close binary companion with a separation of \(0\arcsecond\).67 and a position angle of 321.1$^\circ$, corresponding to a projected separation of 327 au. Due to the proximity the binary system is not astrometrically resolved in the \textit{Gaia} DR3 data.  We derive a prograde, yet misaligned, orbit with $\lambda =41^{+10}_{-9}$\,deg.

\textbf{Qatar-4b} is a massive hot Jupiter. It orbits a supposedly moderately young (170 $\pm$ 10 Myr; but see below) K1 host star on a 1.8-day, possibly slightly eccentric ($e=0.06$), orbit. We find that the orbit of Qatar-4b is well aligned, with $\lambda =-13^{+15}_{-19}$\,deg.

For all the stars considered, the obtained $v\,\sin{i_*}$ values in our R-M effect analysis are in 3-$\sigma$ agreement with the ones reported in the literature from the spectral synthesis. The largest discrepancy is observed for WASP-48. Hence, we have measured the $v\,\sin{i_*}$ from our HARPS-N stacked spectra, using {\tt iSpec} \citep{cua14}. We obtained a value of $v\,\sin{i_*}$ = 11.0 $\pm $ 0.6 km/s, in a better agreement with the value obtained from our R-M effect analysis in Table~\ref{tab:res3}.

The results of our fits are shown in Figs. \ref{f:48}--\ref{f:q4}, while the MCMC results are shown in Figs.~\ref{m48} to \ref{mq4} of the Appendix. The derived values from the MCMC analysis are displayed in Table~\ref{tab:res3}.

\begin{table*}[htbp]
\caption{MCMC analysis results. $\mathcal{N}$ denotes priors with a normal distribution and $\mathcal{U}$ priors with a uniform distribution.}
\small
\label{tab:res3}
\centering
\begin{tabular}{l | c c | c c | c c c} 
\hline\hline
  Parameter & Prior & WASP-48b & Prior & WASP-59b & Prior & WASP-140 Ab \\ 
\hline
$T_c$ - 2450000 [d] & $\mathcal{N}(T_0, 0.006)$ & 8289.542 $\pm$ 0.002 & $\mathcal{N}(T_0, 0.006)$ & 7636.625 $\pm$ 0.003 & $\mathcal{N}(T_0, 0.006)$ & 8394.809 $\pm$ 0.001 \\
$\lambda$ [deg] & $\mathcal{U}(-180, 180)$ & -4 $\pm$ 4 & $\mathcal{U}(-180, 180)$ & -1$^{+20}_{-21}$ & $\mathcal{U}(-180, 180)$ & -1 $\pm$ 3 \\
$v\,\sin{i_*}$ [km/s] & $\mathcal{U}(0, 15)$ & 8.4$^{+1.0}_{-0.9}$ & $\mathcal{U}(0, 15)$ & 0.39$^{+0.13}_{-0.13}$ & $\mathcal{U}(0, 15)$ & 3.0$^{+0.7}_{-0.5}$ \\
$a/R_s$ & $\mathcal{N}(4.66, 0.10)$ & 4.71 $\pm$ 0.08 & $\mathcal{N}(23.58, 0.94)$ & 23.58 $\pm$ 0.93 & $\mathcal{N}(8.58, 0.10)$ & 8.32$^{+0.10}_{-0.10}$ \\
Inc. [deg] & $\mathcal{N}(82.03, 0.5)$ & 81.77 $\pm$ 0.4 & $\mathcal{N}(88.60, 0.19)$ & 88.59$^{+0.19}_{-0.19}$ & $\mathcal{N}(84.3, 0.5)$ & 84.16$^{+0.21}_{-0.20}$ \\
$\Gamma$ [km/s] & $\mathcal{N}(-19.72, 0.20)$ & $-19.718 \pm 0.003$ & $\mathcal{N}(-56.70, 0.20)$ & $-56.699 \pm 0.001$ & $\mathcal{N}(2.19, 0.20)$ & $2.185 \pm 0.001$ \\
$K$ [km/s] & $\mathcal{N}(0.14, 0.20)$ & 0.146 $\pm$ 0.005 & $\mathcal{N}(0.14, 0.20)$ & 0.18 $\pm$ 0.02 & $\mathcal{N}(0.40, 0.20)$ & 0.41 $\pm$ 0.05 \\
\hline
Parameter & Prior & WASP-173 Ab & Prior & TOI-2046b & Prior & HAT-P-41 Ab \\ 
\hline
$T_c$ - 2450000 [d] & $\mathcal{N}(T_0, 0.006)$ & 8410.661$^{+0.001}_{-0.001}$ & $\mathcal{N}(T_0, 0.006)$ & 9852.404$^{+0.002}_{-0.002}$ & $\mathcal{N}(T_0, 0.006)$ & 7575.538$^{+0.001}_{-0.001}$ \\
$\lambda$ [deg] & $\mathcal{U}(-180, 180)$ & 9 $^{+5}_{-5}$ & $\mathcal{U}(-180, 180)$ & -1 $\pm$ 6 & $\mathcal{U}(-180, 180)$ & -4.4 $^{+5.0}_{-5.6}$ \\
$v\,\sin{i_*}$ [km/s] & $\mathcal{U}(0, 10)$ & 6.2$^{+0.55}_{-0.50}$ & $\mathcal{U}(0, 15)$ & 8.9$^{+1.3}_{-1.2}$ & $\mathcal{U}(0, 21)$ & 15.8 $\pm$ 0.8 \\
$a/R_s$ & $\mathcal{N}(4.78,0.17)$ & 4.85$^{+0.14}_{-0.14}$ & $\mathcal{N}(4.75, 0.18)$ & 4.90$^{+0.14}_{-0.14}$ & $\mathcal{N}(5.44,0.10)$ & 5.58$^{+0.11}_{-0.12}$ \\
Inc. [deg] & $\mathcal{N}(85.2, 1.1)$ & 84.78$^{+0.88}_{-0.84}$ & $\mathcal{N}(83.6, 0.9)$ & 82.68$^{+0.71}_{-0.68}$ & $\mathcal{N}(87.7, 1.0)$ & 87.15$^{+1.02}_{-0.82}$ \\
$\Gamma$ [km/s] & $\mathcal{N}(-7.71, 0.20)$ & -7.763$\pm 0.003$ & $\mathcal{N}(-20.64, 0.20)$ & -20.636$\pm 0.007$ & $\mathcal{N}(32.61, 0.20)$ & 32.618$\pm 0.005$ \\
$K$ [km/s] & $\mathcal{N}(0.64, 0.20)$ & $0.612 \pm 0.005$ & $\mathcal{N}(0.37, 0.20)$ & $0.36 \pm 0.02$ & $\mathcal{N}(0.19, 0.20)$ & $0.19 \pm 0.02$ \\

\hline
Parameter & Prior & HAT-P-50b & Prior & Qatar-4b \\ 
\hline
$T_c$ - 2450000 [d] & $\mathcal{N}(T_0, 0.006)$ & 7737.6514$^{+0.003}_{-0.003}$ & $\mathcal{N}(T_0, 0.006)$ & 8058.423$^{+0.002}_{-0.002}$ \\
$\lambda$ [deg] & $\mathcal{U}(-180, 180)$ & 41$^{+10}_{-9}$ & $\mathcal{U}(-180, 180)$ & -13 $^{+15}_{-19}$ \\
$v\,\sin{i_*}$ [km/s] & $\mathcal{U}(0, 15)$ & 8.5$^{+1.5}_{-1.4}$ & $\mathcal{U}(0, 15)$ & 5.1$^{+1.0}_{-0.9}$ \\
$a/R_s$ & $\mathcal{N}(5.68, 0.19)$ & 5.61$^{+0.18}_{-0.19}$ & $\mathcal{N}(7.1, 0.5)$ & 7.25$^{+0.43}_{-0.40}$ \\
Inc. [deg] & $\mathcal{N}(83.65, 0.57)$ & 83.86$^{+0.53}_{-0.50}$ & $\mathcal{N}(87.5, 1.6)$ & 87.45$^{+1.68}_{-1.42}$ \\
$\Gamma$ [km/s] & $\mathcal{N}(6.48, 0.20)$ & 6.448$\pm 0.005$ & $\mathcal{N}(-28.74, 0.20)$ & -28.751$\pm 0.008$ \\
$K$ [km/s] & $\mathcal{N}(0.20, 0.20)$ & 0.18 $\pm 0.03$ & $\mathcal{N}(1.00, 0.20)$ & 0.94 $\pm$ 0.03 \\
\hline
\end{tabular}
\end{table*}

\section{Discussion}
\subsection{Stellar rotation and true obliquity $\psi$ }

By detecting variations in the light curve, we can determine the stellar rotation periods \citep[e.g.,][]{ska22}. This is crucial for estimating the stellar inclination $i_*$, which in turn helps to determine the true obliquity. We attempted to derive the stellar rotation periods by analyzing ASAS-SN \textit{g}-band light curves \citep{ko17} using a Lomb-Scargle periodogram \citep{lomb76,sca82}. We were able to detect a periodic modulation of WASP-140 A with a period of $10.44 \pm 0.05$ days (with False Alarm Probability (FAP)=$1.6\times10^{-11}$, Fig.\,\ref{fig:perrot1}), in agreement with the rotational period of $10.4 \pm 0.1$ days reported by \citet{hel17} using Wide Angle Search for Planets (WASP) survey data. For Qatar-4, we detect a period of $7.07 \pm 0.08$ days (FAP=$1.6\times10^{-9}$, Fig.\,\ref{fig:perrot2}). For WASP-173 A, we detect a period of $7.97 \pm 0.05$ days (FAP=$1.6\times10^{-12}$, Fig.\,\ref{fig:perrot3}), also in agreement with the rotational period of $7.9 \pm 0.1$ days reported by \citet{hel19} from WASP data. Furthermore, we have explored \textit{TESS} data \citep{rick15} in search of rotational modulation. TOI-2046 was observed in 7 sectors. We detect a periodic modulation of TOI-2046 with a period of $4.05 \pm 0.05$ days (FAP=$10\times10^{-9}$, Fig.\,\ref{fig:perrot4}) from the long cadence data in 3 sectors and a period of $4.29 \pm 0.05$ days (FAP=$10\times10^{-4}$) using the short cadence data in four different sectors. Such variability is present in many F-type stars showing rotational variability \citep[e.g.,][]{hen23}. Due to the higher SNR detection, we adopt the former period. This periodicity, that we assume to be due to differential rotation, was not detected in the ASAS-SN data due to the small amplitude. For the other targets, we were not able to detect any signal with high significance.

From the derived rotational periods we derive the equatorial velocity $v_{eq}$ of $4.2 \pm 0.2$ km/s for WASP-140 A, $7.0 \pm 0.4$ km/s for WASP-173 A, $15.1 \pm 1.1$ km/s for TOI-2046 and $6.1 \pm 0.5$ km/s for Qatar-4. These values are in good agreement with the $v\,\sin{i_*}$ values reported in Tables~ \ref{tab:targets1} and \ref{tab:res3}. The different methods of deriving the rotational velocities might produce slight variations as, for example, the spectral synthesis method might not be able to distinguish various broadening mechanisms as opposed to the values derived from the R-M effect analysis that should provide only the rotational velocity. Another origin of the small variations can be a differential rotation of the star. \citet{rei13} found that the differential rotation increases with effective temperature and above 6000 K can be quite scattered. For example, if a planet is transiting over high stellar latitudes, the derived $v\,\sin{i_*}$ may be lower compared to the rotational velocity originating from stellar spots in lower latitudes where the rotation is faster. Finally, as thoroughly discussed in \citet{bro17}, the Boué model \citep{bou13} for the Rossiter-McLaughlin effect, which we employed and that was developed to be used with instruments like HARPS, is known to provide a lower value of $v\,\sin{i_*}$ compared to other models such as the Hirano model \citep{hir11} developed for the iodine cell or the Doppler shadow method \citep{coll10}, yet they agree on the derived projected obliquity value.

For these four planets, we are thus able to infer the true stellar obliquity $\psi$ (i.e., the angle between the stellar spin-axis and the normal to the orbital plane) using the spherical law of cosines, $\cos \psi = \sin i_*\,\sin i\,\cos |\lambda|+\cos i_*\,\cos i$ with the approach suggested by \citet{mas20} accounting for the dependency between $v$ and $v\,\sin{i_*}$. We obtain $\psi = $30$^{+18}_{-15}$\,deg for WASP-140 Ab, $\psi = $21$^{+9}_{-10}$\,deg for WASP-173 Ab, $\psi =$42$^{+10}_{-8}$\,deg for TOI-2046b and $\psi = $32$^{+14}_{-13}$\,deg for Qatar-4b. From the true obliquity measurements, TOI-2046b shows a rather misaligned orbit. For the rest of the targets we can conclude that the orbits are clearly prograde, yet not exactly aligned with their host stars, although the misalignment that we measure is not statistically significant. In the following sections, we discuss the tidal alignment timescale and how the R-M effect and atmospheric characterization are in high synergy to unravel the history of the system.

\subsection{Obliquity of young systems}

The stellar obliquities of young systems are still not well characterized. There are only a handful of obliquity measurements for systems with ages below 200 Myr. These young systems offer a unique opportunity to study the primordial misalignment as, generally, the tidal forces have not had enough time to alter the architecture of the system. Currently, all the planets younger than 200 Myr seem to be well aligned with their host stars \citep{alb22}. Qatar-4 with a reported age of 170 $\pm$ 10 Myr \citep{als17} and TOI-2046 with its reported age between 100-400 Myr \citep{kab22} would have been valuable additions to the few known systems. In Appendix~\ref{age}, we determine the age of Qatar-4 and TOI-2046 with several methods (lithium depletion, gyrochronology and $ R'_{HK} \, \text{index} $) and we provide evidence for Qatar-4 to be at least 350-500 Myr old and TOI-2046 being at least 700 Myr old. Hence, they do not belong to the moderately young population of exoplanets below 200 Myr.

\subsection{Dynamical timescale}

To better understand the evolution of the studied systems and their relevance, here we compare the age of the system and the tidal alignment timescale that can be used to study whether the spin-orbit angle has changed since the formation of the system. It provides only a limit on the possible realignment as the spin-orbit misalignment does not have to happen at the same time as the system's formation. For cool stars below the Kraft break \citep[][that is, with convective envelopes]{kraft67}, the tidal alignment timescale can be approximated \citep{alb12} as

\begin{equation}
\label{ce}
\hspace*{0.25\columnwidth}    \tau_{\rm{CE}} = \frac{10^{10}\,\rm{yr}}{\left( M_{\rm{p}}/M_{\rm{*}} \right)^2}  \left( \frac{a/R_{\rm{*}}}{40} \right)^6.
\end{equation}

\noindent
This allows us to compute the tidal alignment timescale for all the studied planets but HAT-P-41 Ab, and derive the following timescales: $\tau_{\rm{Ce}} = 4 \cdot 10^{10}$\,yr for WASP-48b, $4 \cdot 10^{14}$\,yr for WASP-59b, $1 \cdot 10^{11}$\,yr for WASP-140 Ab, $3 \cdot 10^{9}$\,yr for WASP-173 Ab, $5 \cdot 10^{10}$\,yr for TOI-2046b, $8 \cdot 10^{10}$\,yr for HAT-P-50b and $1 \cdot 10^{10}$\,yr for Qatar-4b.

HAT-P-41 A, with its spectral type F6, is classified as a hot star and thus lies above the Kraft break,  having a radiative envelope -- this structural difference has implications on the tidal forces and a different formula of the tidal alignment timescale \citep{alb12} should be used:

\begin{equation}
\label{ra}
\hspace*{0.08\columnwidth}    \tau_{\rm{RA}} = \frac{5}{4} \cdot 10^9\, \text{yr}  \left( \frac{M_{\rm{p}}}{M_{\rm{*}}} \right)^{-2}  \left(1 + \frac{M_{\rm{p}}}{M_{\rm{*}}}\right)^{-5/6}  \left( \frac{a / R_{\star}}{6} \right)^{17/2},
\end{equation}

\noindent
leading to $\tau_{\rm{RA}} = 2 \cdot 10^{15}$\,yr for HAT-P-41 Ab.

Due to its short orbital period and large mass, WASP-173 Ab is likely older than its tidal alignment timescale and, hence, the high-eccentricity migration scenario cannot be ruled out from the stellar obliquity measurements alone. From the derived tidal alignment timescales of the rest of the planets, we can conclude that the current spin-orbit angles were likely not significantly changed from their initial values through tidal forces.

The only misaligned planet in our sample is HAT-P-50b. Given the reported presence of a close companion, this would be consistent with high-eccentricity migration via the Kozai-Lidov mechanism \citep{fab07}. We have therefore estimated the Kozai-Lidov timescale $\tau_{\rm{KL}}$ to check whether such a scenario is viable, using the following formula \citep{naoz16}:

\begin{equation}
\hspace*{0.18\columnwidth} \tau_{\text{KL}} \approx \frac{16}{30\pi} \frac{P_{\text{B}}^2}{P_{\text{b}}} (1 - e_{\text{B}}^2)^{3/2} \left(\frac{M_{\text{A}} + M_{\text{B}}}{M_{\text{B}}}\right),
\end{equation}

where $M_{\text{A}}$ and $M_{\text{B}}$ are the masses of the primary and secondary components of the binary system, $P_{\text{B}}$ is the orbital period of the binary system, $P_{\text{b}}$ the orbital period of the exoplanet and $e_{\text{B}}$ the eccentricity of the binary pair. We have used values of $M_{\text{B}}$=0.62 M$_\odot$ and $P_{\text{B}}$=4\,440 yrs \citep{les21}. As the eccentricity of the binary system remains unconstrained, we calculate the $\tau_{\rm{KL}}$ for values of $e_{\text{B}}$ between 0 to 0.9, leading to $\tau_{\rm{KL}} \sim 1$\,Gyr for $e_{\text{B}}$=0 and $\tau_{\rm{KL}} \sim 0.1$\,Gyr for $e_{\text{B}}$=0.9. From the derived values and their comparison to the age of the system of $3.37^{+1.44}_{-0.27}$\,Gyr \citep{hart15}, we find that the Kozai-Lidov mechanism has possibly had enough time to change the orbital configuration of the system. Additionally, this timescale would be even shorter in the likely scenario that the planet has formed on a wider than currently observed orbit. To further confirm this orbital evolution hypothesis, measurements of the planetary atmospheres are warranted \citep{pen24}, but as shown in Table~\ref{tab:met} in the Appendix, HAT-P-50b is rather not an amenable target for atmospheric studies. 

The low values of the spin-orbit angle of the remaining planets are consistent with slow disk migration \citep{bar14}. Planets undergoing disc migration are expected to have a distinct composition to those arriving via high-eccentricity migration \citep[][]{kirk24} and we comment on the future atmospheric prospects of HAT-P-41 Ab in the section \ref{atmo} of the Appendix. 

\begin{figure*}
\includegraphics[width=\textwidth]{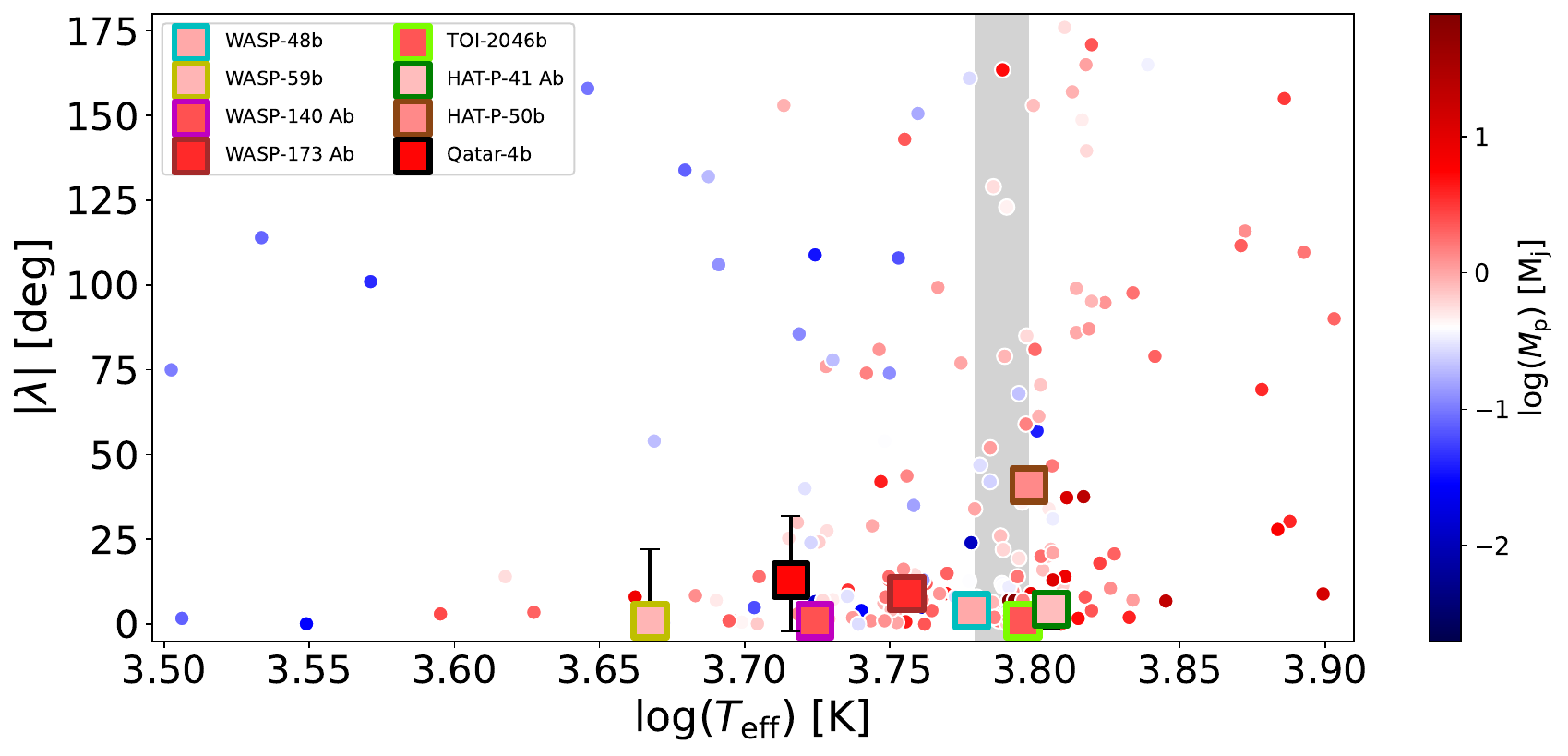}
\caption{Projected obliquity versus stellar effective temperature. Our targets are plotted with squares. The gray area shows the position of the Kraft break as derived in \citet{spal22}. The literature values were retrieved from the TEPCat catalogue \citep{south11}.}
\label{dist}
\end{figure*}

We show the comparison with the rest of the population in Fig. \ref{dist}. HAT-P-50b lies in the less populated region of Hot Jupiters with misaligned orbits. These planets are important pieces in unraveling whether misaligned planets are distributed isotropically as suggested by \citet{dong23} and \citet{sie23} or whether the misaligned population shows a preference for polar and retrograde orbits as suggested by \citet{alb21} and \citet{att23}. Furthermore, HAT-P-50 lies on the position of the metallicity-dependent Kraft break as derived by \citet{spal22} where convective envelopes persist in a lower metallicity environment for slightly higher temperatures. The rest of our targets join the large population of Hot Jupiters on prograde and aligned orbits around their host stars.

As discussed in \citet{hel17}, WASP-140 Ab has a very short circularisation timescale, $\tau_{\mathrm{cir}}$, estimated to be between $\sim$1 and 100 Myr (for a quality factor $Q_p = 10^5$ and $10^6$, respectively), hence a circular orbit is expected. However, the system retains a small but non-zero eccentricity. A recent migration event could explain the non-circular orbit, together with $\psi = 21^{+9}_{-10}$ derived above. Another plausible explanation is that an outer planetary companion in the WASP-140 A system is dynamically maintaining the non-zero eccentricity. This hypothesis makes the WASP-140 A system an excellent target for further investigation to resolve this ambiguity.

\subsection{Obliquity in multi-star systems}

The spin-orbit alignment is being investigated in multi-star systems to understand how planets form and evolve in binary systems and how the presence of the binary companion affects the proto-planetary disk environment \citep{chris24}. This can be done by investigating the orbital orientation between exoplanets and wide-orbiting binary companions. \citet{chris24} reported the existence of an alignment between the orbits of small planets and binary systems with semimajor axes below 700 au. \citet{rice24} have studied the obliquity distribution of multi-star systems. They have reported no clear correlation between spin-orbit misalignment and orbit-orbit misalignment. Furthermore, they report the discovery of an overabundance of systems that are consistent with joint orbit-orbit and spin-orbit alignment.

In the sample presented here, three planets are in astrometrically resolved binary systems: WASP-140, WASP-173 and HAT-P-41. This allows us to compute the orbit-orbit angle $\gamma$ using \textit{Gaia} DR3 data:

\begin{equation}
\hspace*{0.35\columnwidth} \cos \gamma = \frac{\vv{r} \cdot \vv{v}}{|\vv{r}| |\vv{v}|},
\end{equation}

where $\vv{r} \equiv [\Delta \alpha, \Delta \delta]$ and $\vv{v} \equiv [\Delta \mu^*_{\alpha}, \Delta \mu_{\delta}]$. Here, $\Delta \alpha$ and $\Delta \delta$ denote the positional differences between the primary and secondary in the right ascension (RA) and declination (Dec) directions, while $\Delta \mu^*_{\alpha}$ and $\Delta \mu_{\delta}$ are the proper motion differences in the RA and Dec directions. \textit{Gaia} DR3 reports a $\Delta \mu^*_{\alpha}$ value that has already the $ \cos \delta$ corrective factor in the RA direction incorporated\footnote{See the pmra definition in the Gaia DR3 source documentation at \url{https://gea.esac.esa.int/archive/documentation/GDR3/Gaia_archive/chap_datamodel/sec_dm_main_source_catalogue/ssec_dm_gaia_source.html}.} \citep{rice24}.

The angle \( \gamma \) quantifies the alignment between a binary system's relative position vector, $\vv{r}$, and relative velocity vector, $\vv{v}$. Values of \( \gamma \) near \( 0^\circ \) or \( 180^\circ \) suggest that the binary's motion is predominantly along our line of sight, characteristic of an edge-on orientation (orbit-orbit alignment). Conversely, \( \gamma \) values near \( 90^\circ \) indicate motion primarily perpendicular to our line of sight, aligning with a face-on orientation (orbit-orbit misalignment). However, interpreting \( \gamma \) for individual systems involves complexities: orbital eccentricity can cause variations in $\vv{r}$ and $\vv{v}$ that affect \( \gamma \). Furthermore, the planet-hosting star can be oriented in within the sky-plane hence accurately determining the alignment between the planetary orbit and the binary orbit is difficult. Therefore, while \( \gamma \) offers insights into the system's geometry, caution is necessary when interpreting its value due to these potential degeneracies. Significant improvements will be brought by the next \textit{Gaia} data releases as the current astrometry data are based solely on 34 months of observations \citep{gaia23}.

For WASP-140 Ab, we infer a well-aligned spin-orbit angle of $\lambda =-1 \pm 3$\,deg. Furthermore, we derive an almost face-on orbit-orbit orientation with $\gamma =80.3 \pm 6.3$\,deg and a binary inclination of $78.8 \pm 5.9$\,deg. For WASP-173 Ab we derive a well-aligned orbit with $\lambda =9 \pm 5$\,deg; additionally, we derive the orbit-orbit orientation of $\gamma =136.8 \pm 4.2$\,deg and a binary inclination of $65.1 \pm 7.0$\,deg. Finally, our R-M effect result ($\lambda =-4.4^{+5.0}_{-5.6}$\,deg) puts HAT-P-41 into the group of systems with spin-orbit alignment and orbit-orbit misalignment ($\gamma =72.7 \pm 2.9$\,deg).

Since the study by \citet{rice24}, two more systems in multi-star systems had their obliquity measured (WASP-77 and HD 110067). WASP-77 is a binary system with 1.00 $\pm$ 0.07 M$_\odot$ and 0.71 $\pm$ 0.06 M$_\odot$ components. The primary star hosts a hot Jupiter on a 1.4-day orbit \citep{max13}. \citet{zak24a} reported $\lambda =-8^{+19}_{-18}$\,deg for WASP-77 Ab. Using Gaia DR3 astrometry \citep{gaia23}, we can now also compute the orbit-orbit alignment, $\gamma =14.4 \pm 12.1$\,deg, together with the binary inclination of $86.5 \pm 15.1$\,deg. Hence, the WASP-77 system joins 8 other systems identified by \citet{rice24} that also show such a high degree of alignment. HD 110067 hosts six transiting sub-Neptunes in resonant chain orbits. The primary has an equal mass binary companion at separation of 13\,400 au making this a triple system with the highest number of known exoplanets \citep{luq23,app23}. \citet{zak24b} inferred an aligned orbit for HD 110067c with $\lambda =6^{+24}_{-26}$\,deg. We now report also the orbit-orbit alignment of $\gamma =156.5 \pm 23.2$\,deg, hinting at a possible slight deviation from a fully edge-on orbit.

\section{Summary}
Understanding the spin-orbit orientation in compact systems and how planetary orbits align or misalign with the orbits of multi-star systems and what pathways these planets might follow to achieve their observed configuration (e.g., the reported preference for line-of-sight orbit-orbit alignment) still remains unanswered. To improve our understanding of these pathways we have studied the projected stellar obliquity of eight gas giants on short orbits around FGK stars with five systems being part of multi-star systems. From the archival HARPS and HARPS-N data, we have measured the Rossiter-McLaughlin effect to infer the projected spin-orbit alignment of WASP-40b, WASP-59b, WASP-140 Ab, WASP-173 Ab, TOI-2046b, HAT-P-50b and Qatar-4b for the first time and we have updated the value for HAT-P-41 Ab using a new dataset. We report a prograde but misaligned orbit for HAT-P-50b ($\lambda =41^{+10}_{-9}$\,deg) consistent with a Kozai-Lidov mechanism triggered by the close stellar companion. We derive potentially aligned orbits for the rest of the studied systems ruling out highly excited orbits due to the presence of stellar companions. For four systems, we have also derived their true (3-D) obliquity with TOI-2046b having a misaligned orbit with $\psi =$42$^{+10}_{-8}$\,deg. Additionally, we derive $\psi = $30$^{+18}_{-15}$\,deg for WASP-140 Ab, $\psi = $21$^{+9}_{-10}$\,deg for WASP-173 Ab, for and $\psi = $32$^{+14}_{-13}$\,deg for Qatar-4b.
We provide a refined age estimate for the previously reported moderately young ($100$ to $400$ Myr) system TOI-2046 and ($\sim 170$ Myr) system Qatar-4; we rather conclude that the systems are at least 700 Myr and 350-500 Myr old, respectively. Finally, we report the orbit-orbit angle for five multi-star systems hosting exoplanets for the first time identifying two systems with line-of-sight orbit-orbit misalignment. Further studies of the spin-orbit orientation in multi-star systems are highly encouraged especially with the upcoming \textit{Gaia} DR4 release coming in 2026 providing astrometry data based on 66 months of observations to enhance the orbit-orbit constraints.

\begin{acknowledgements} The authors would like to thank the anonymous referee for
their insightful report. JZ and PK acknowledge support from GACR:22-30516K. AB is supported by the Italian Space Agency (ASI) with Ariel grant n. 2021.5.HH.0. MS acknowledges the financial support of the Inter-transfer grant n. LTT-20015.
\end{acknowledgements}

\bibliographystyle{aa}
\bibliography{aa}

\appendix

\section{TESS photometry of WASP-59}
\label{sec:w59photo}

We analyzed photometric data of WASP-59 (TIC 91051152) using observations from \textit{TESS} \citep{rick15}. The target was observed in a single Sector 56, during September 2022 at a 120-second cadence. A total of three transits of WASP-59b were captured, which were used to improve the planetary transit parameters.
We utilized the PDCSAP flux, which provides light curves corrected for instrumental systematics and stellar variability. To model the observed transits, we employed the \texttt{exoplanet} package \citep{foreman21}, a Bayesian framework for transit modeling.

\begin{figure}[h]
\includegraphics[width=0.5\textwidth]{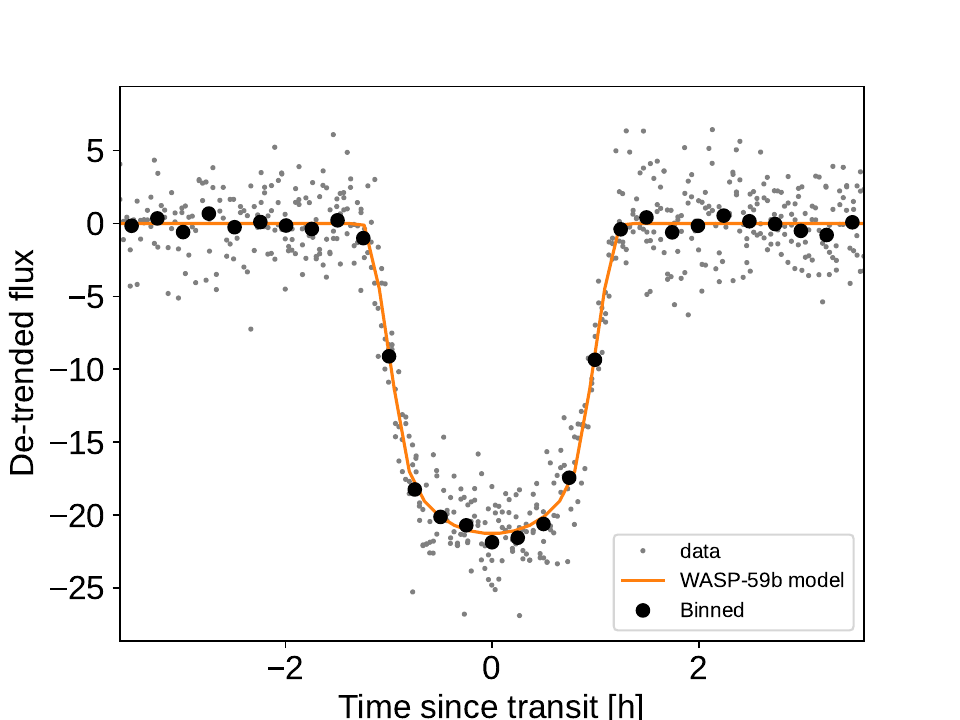}
\caption{Phase curve of WASP-59b using \textit{TESS} data centered around the primary transit. Observed data are shown in gray, binned data in black and the obtained transit model in orange color.}
\label{fig:w59photo}
\end{figure}

The transit model assumed a Keplerian orbit and included a Gaussian Process (GP) to account for residual stellar variability. The stellar radius ($R_\star$) was constrained with a Gaussian prior centered on the value reported by \citet{heb13}. Similarly, the orbital period ($P$) was modeled using a Gaussian prior based on previously determined ephemeris values. The transit depth ($\delta$) was also constrained with a Gaussian prior informed by earlier observations. A uniform prior between 0 and 1 was applied to the impact parameter ($b$). We adopted a quadratic limb-darkening law following \citet{kipping13}. The limb-darkening coefficients ($u_1, u_2$) were sampled with uniform priors constrained to the range between 0 and 1. The eccentricity $e$ and argument of periastron $\omega$ were fixed to values from \citet{heb13}. To model residual stellar variability in the PDCSAP flux, we used a Matern-3/2 Gaussian Process kernel \citep{foreman17, foreman18}. We initialized the model using estimates for the orbital period, planetary radius, and reference transit time obtained from a Box Least Squares (BLS) periodogram analysis. The model parameters were first optimized using the maximum a posteriori (MAP) estimate. This was followed by Markov Chain Monte Carlo (MCMC) sampling using the No-U-Turn Sampler (NUTS). We performed 2000 tuning steps and sampled the posterior distributions with 7000 draws using three independent ensembles ensuring R-hat $\hat{R}$ value below 1.001 for all parameters. The obtained results are in good agreement with the literature values and are shown in Table \ref{tab:w59photo} and the phase curve shown in Fig.~\ref{fig:w59photo}.

\begin{table}
\centering
\caption{Obtained parameters of WASP-59b from fitting \textit{TESS} data.}
\label{tab:w59photo}
\begin{tabular}{l c}
\hline
\hline
\textbf{Parameter} & \textbf{Value} \\
\hline
\vspace{0.05cm}
$T_0$ (TJD) & $2830.348646^{+0.000407}_{-0.000400}$ \\
\vspace{0.05cm}
$P$ (days) & $7.919567^{+0.000079}_{-0.000079}$ \\
\vspace{0.05cm}
$u_{\star, 0}$ & $0.32^{+0.20}_{-0.20}$ \\
\vspace{0.05cm}
$u_{\star, 1}$ & $0.33^{+0.34}_{-0.34}$ \\
\vspace{0.05cm}
$a / R_\star$ & $23.58^{+0.94}_{-0.91}$ \\
\vspace{0.05cm}
R$_{\rm{p}}$/R$_{\rm{s}}$ & $0.1389^{+0.0030}_{-0.0032}$ \\
\vspace{0.05cm}
$i$ (deg) & $88.60^{+0.19}_{-0.19}$ \\
\hline
\end{tabular}
\end{table}

\section{Doppler Shadow of HAT-P-41 Ab}
\label{DS}

\begin{figure*}[h]
\includegraphics[width=\textwidth]{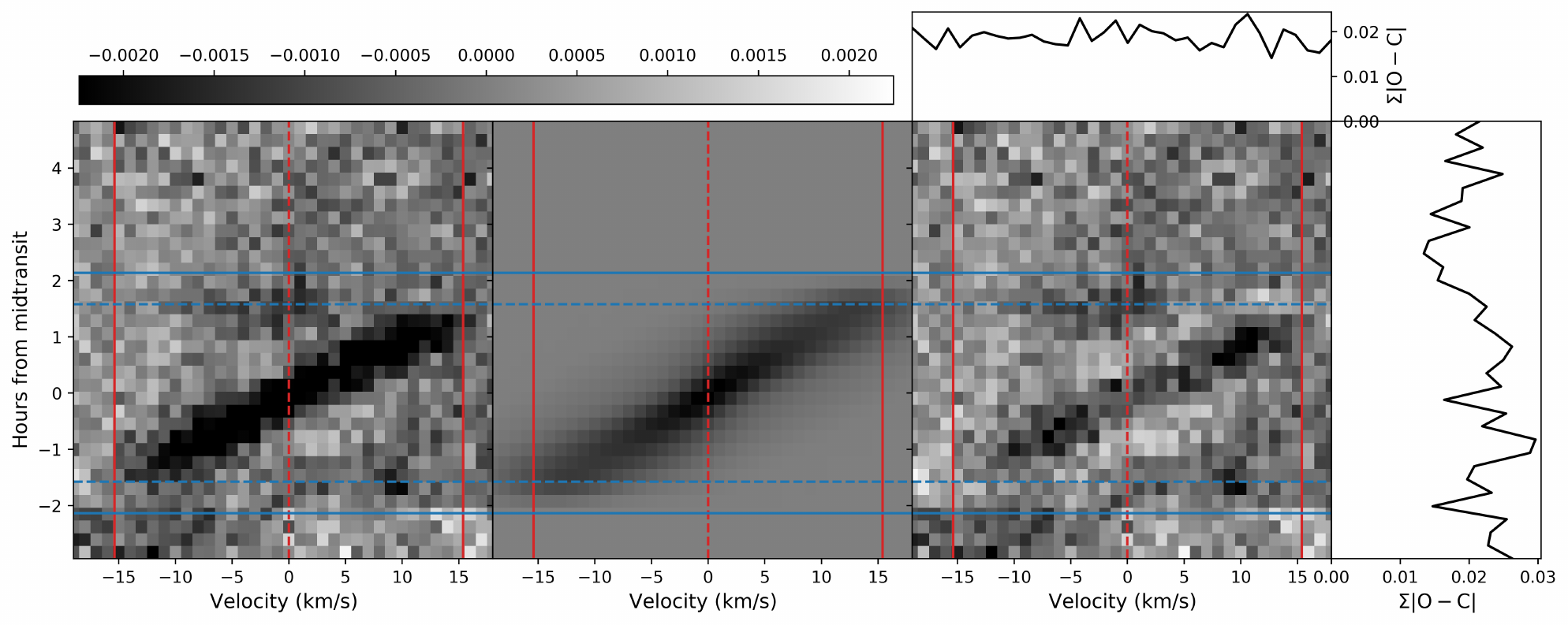}
\caption{Left: Doppler shadow of HAT-P-41 Ab as it appears in the observed data (depth is indicated by the upper, horizontal colorbar); middle: a best-fitting model of the Doppler shadow, right: residuals of the best-fitting model after subtraction from the actual data. The blue solid and dashed lines mark the boundaries for the full and total transit times, respectively.}
\label{fig:DS}
\end{figure*}

Doppler tomography, a technique \citep{coll10} used to measure the stellar obliquity independently on the classical method provides a high-fidelity representation of the integrated stellar line profile, which is essential for detecting distortions induced during a planetary transit. By subtracting individual in-transit CCFs from a master out-of-transit CCF, which represents the underlying Gaussian-approximated line shape, one can isolate the characteristic time-varying distortion, known as the "Doppler shadow." This shadow encodes valuable information about the projected rotational velocity of the stellar surface ($v\,\sin{i_*}$) and the sky-projected obliquity ($\lambda$). We analyzed HARPS-N data of HAT-P-41. Using the Doppler tomography method, we successfully detected the Doppler shadow, as shown in Figure \ref{fig:DS}. To fit the shadow of HAT-P-41 Ab, we employed \textsc{tracit}, a Python module\footnote{\url{https://tracit.readthedocs.io/}} developed and described in \cite{knu22} that implements MCMC method.

\begin{figure}[h]
\includegraphics[width=0.5\textwidth]{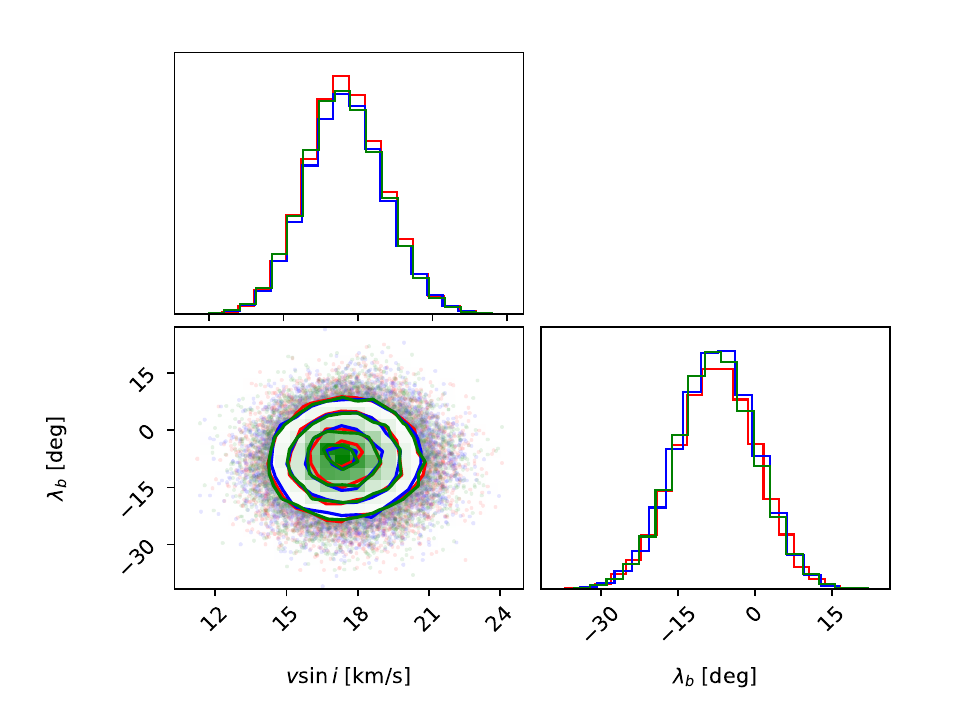}
\caption{Corner plot representing the posterior distributions of the projected rotational velocity ($v\,\sin{i_*}$) and projected obliquity ($\lambda$) of our Doppler Shadow analysis of HAT-P-41 Ab. Shown are three MCMC initializations. The result ($\lambda_b$ = -7.6$^{+7.7}_{-8.0}$ deg) prefers an aligned orbit that is consistent with our classical R-M effect analysis.}
\label{fig:DSres}
\end{figure}

Setting a uniform uninformative prior on $\lambda$ did not allow us to meaningfully distinguish between the results from \citet{john17} and the results we have previously derived. Hence, we have set a Gaussian prior on $\lambda$ to -13.1 deg as the mean value between the $\lambda$ from \citet{john17}, and the one that we previously derived using the classical RM effect. The same approach was applied for $v\,\sin{i_*}$, where we set the central value 17.7 km/s (the mean of 19.6 and 15.8 km/s), coming from mentioned sources, respectively. We ran three independent MCMC realizations using 100 walkers with 12000 draws, burning 6000 with the corner plot shown in Fig. \ref{fig:DSres}. The priors we used are listed in Table \ref{tab:priors}, together with the obtained results. The used values for microturbulence $\zeta$, and macroturbulence $\xi$ velocities utilized were taken from \citet{bruntt10} and \citet{doy14}, respectively. The limb darkening values $q_1$ and $q_2$ were obtained from the ExoCTK tool. The Gelman-Rubin statistic for each parameter was below 1.004. The obtained result of $\lambda_b$ = -7.6$^{+7.7}_{-8.0}$ deg favors an aligned orbit of the planet in agreement with our classical R-M effect analysis.

\begin{table}[h!]
\centering
\caption{Priors and results of the HAT-P-41\,Ab Doppler shadow analysis.}
\begin{tabular}{lccc}
\hline
Parameter & Priors & Result \\
\hline
$\lambda_b$ [deg] & $\mathcal{N}(-13.1, 8.7)$ & -7.6$^{+7.7}_{-8.0}$    \\
\vspace{0.05cm}
$v\sin i$ [km/s] & $\mathcal{N}(17.7, 1.9)$ & 17.5$^{+1.6}_{-1.5}$ \\
\vspace{0.05cm}
$\xi$ [km/s] & $\mathcal{N}(1.5, 1.0)$ & $4.3^{+1.8}_{-4.0}$ \\
\vspace{0.05cm}
$\zeta$ [km/s] & $\mathcal{N}(6.0, 1.0)$ & $7^{+2}_{-3}$ \\
\vspace{0.05cm}
$q_1 + q_2$ & $\mathcal{N}(0.657, 0.1)$ & $0.67^{+0.1}_{-0.1}$ \\
\hline
\end{tabular}
\label{tab:priors}

\end{table} 

\section{Stellar ages}
\label{age}
\begin{figure*}
\includegraphics[width=1.0\textwidth]{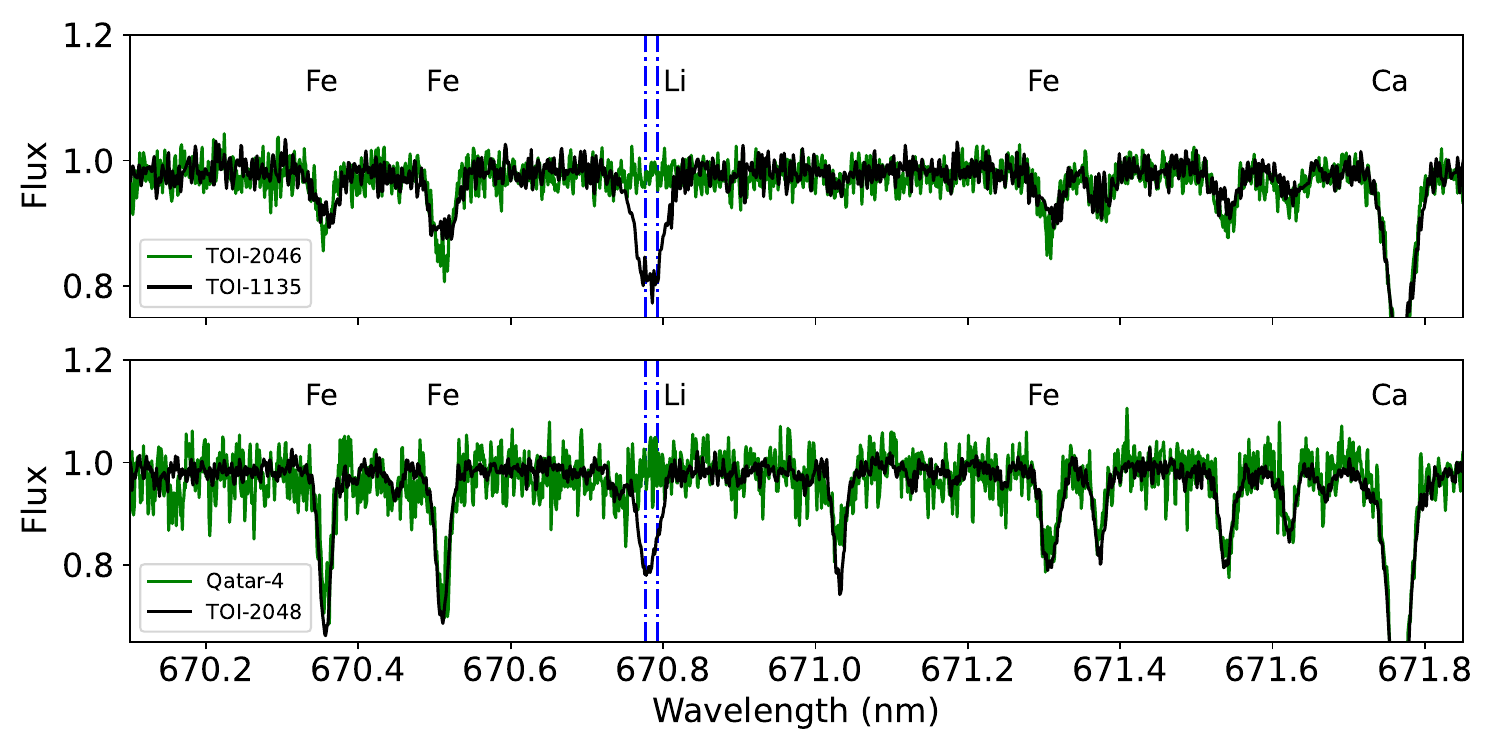}
\caption{Spectral region around the lithium doublet (blue vertical lines) together with iron and calcium lines. The top panel shows stacked spectra of TOI-2046 (green) alongside with TOI-1135 (black) with similar spectral type and an age $\sim$ 650 Myr \citep{mall24}. The bottom panel shows stacked spectra of Qatar-4 (green) alongside with TOI-2048 (black) with similar spectral type and an age $\sim$ 300 Myr \citep{new22}. Both TOI-2046 and Qatar-4 show clear non-detection of the lithium hinting at an older age.} 
\label{fig:li}
\end{figure*}

We have tried to improve the age determination of the supposedly young stars TOI-2046 and Qatar-4 using the lithium method. Despite increasing the signal by stacking all the HARPS-N spectra, we report a non-detection of lithium in the stellar spectrum of both TOI-2046 and Qatar-4 (Fig.\,\ref{fig:li}). Hence, we can only derive lower age limits: using lithium models from \citet{jeff23} for the given effective temperature of TOI-2046 and Qatar-4, we infer that TOI-2046 is at least 700 Myr old and Qatar-4 is at least 500 Myr old. 

Additionally, we have applied the $ R'_{HK} \, \text{index} $ method. We have measured the median value of $\text{log}\,R'_{HK} = $ -4.66 $\pm$ 0.01 for TOI-2046 and $\text{log}\,R'_{HK} =$ -4.53 $\pm$ 0.02 for Qatar-4. Using the calibrations from \citet{mam08} we infer activity age of 1.6 $\pm$ 0.1 Gyr for TOI-2046 and 0.75 $\pm$ 0.10 Gyr for Qatar-4.

\begin{figure}[h!]
\includegraphics[width=0.5\textwidth]{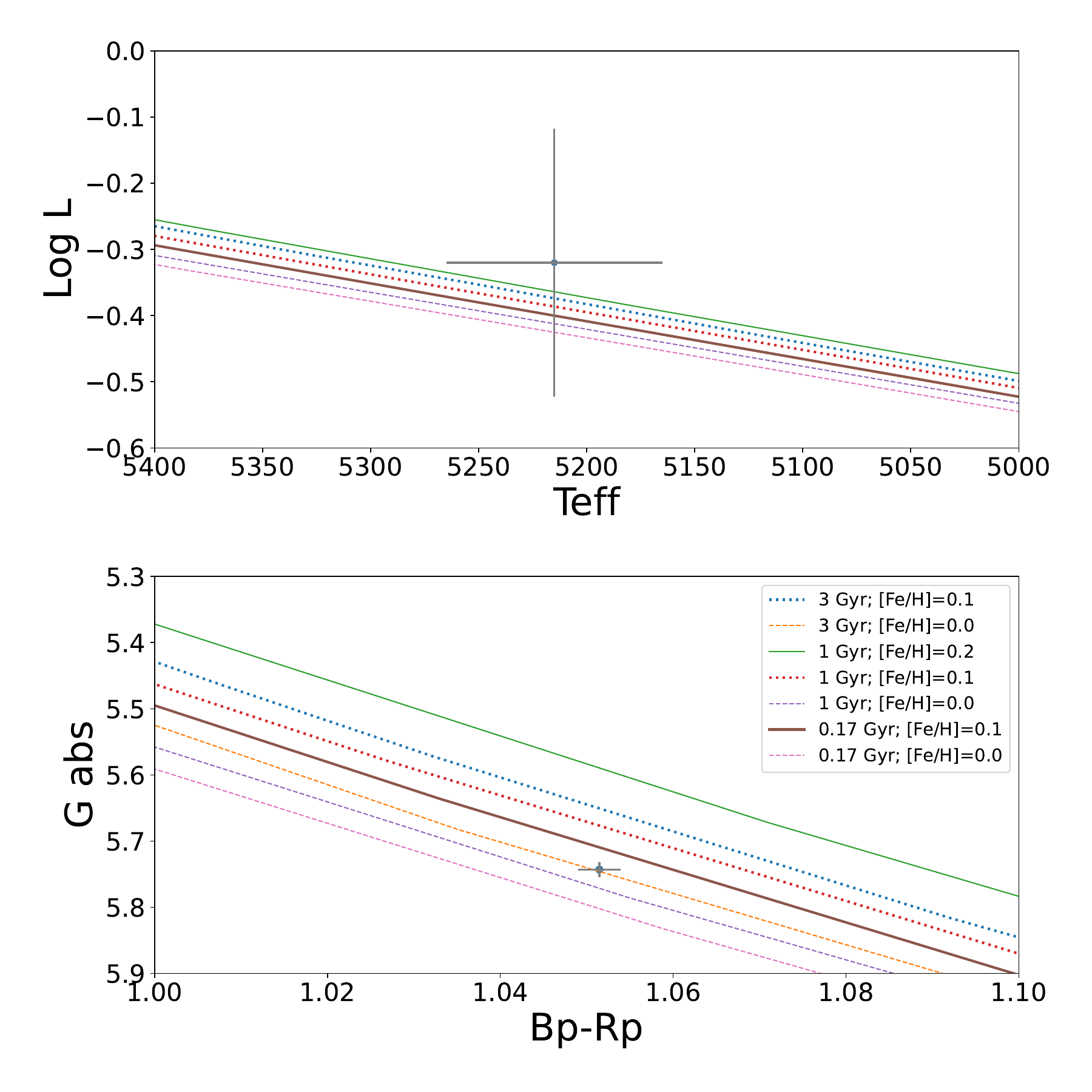}
\caption{Top: H-R diagram, using effective temperature and luminosity of Qatar-4. Bottom: \textit{Gaia} color-magnitude diagram. In both cases, the position of Qatar-4 is indicated, while the colored lines show distinct isochrones, as indicated in the legend. This clearly indicates that isochrones only cannot be used to derive a precise age.}
\label{isoq4}
\end{figure}

Subsequently, we have tried to infer the age of the Qatar-4 using isochrones. From the stacked spectra we have determined the stellar parameters of Qatar-4. We have used the iSpec software \citep{blanco14,blanco19} together with MARCS atmosphere models \citep{gusta08}. We derive T$_{\rm{eff}}$ =5220 $\pm$ 45 K, $\log g$ = 4.49 $\pm$ 0.09, [Fe/H]= 0.09 $\pm$ 0.10 and $v\,\rm{sin}i_*$ = 5.7 $\pm$ 0.3 km/s. The derived parameters are in agreement with previous work using TRES spectra \citep{als17}. Using the above-derived parameters as well as the Gaia magnitudes, we were not able to put any meaningful constraint on the age from only the isochrones fitting (Fig. \ref{isoq4}).

We have also determined the age of Qatar-4 using gyrochronology. The gyrochronology method employs the age-rotation relation to infer the ages of stars. Such a method is based on the observations of stellar clusters that showed that stellar rotation slows down as the stars become older. First, we have checked that the RUWE value in the Gaia DR3 catalog is not above 1.25, which would imply a poor astrometric solution possibly hinting at an unresolved binary companion that could bias subsequent analysis. The retrieved RUWE value of 1.02 points towards a well-behaved astrometric solution. We have used the python package \texttt{gyro-interp}\footnote{\url{https://github.com/lgbouma/gyro-interp/}} based on the work of \citet{bouma23}. For the stellar rotational period and the effective temperature of Qatar-4, we derive an age of 350$^{+120}_{-111}$ Myr. As an independent check, we have also used the \texttt{stardate}\footnote{\url{https://github.com/RuthAngus/stardate/}} python tool that combines isochrone fitting with gyrochronology \citep{ang19}. We derive an age of 500$^{+300}_{-150}$ Myr. Hence, Qatar-4 does not appear to belong to the moderately young ($<$ 200 Myr) population of stars hosting exoplanets. Following the same steps, we also derive an age of 0.46$^{+0.25}_{-0.15}$ Gyr for TOI-2046, using the \texttt{gyro-interp} tool, and 0.8$^{+1.1}_{-0.5}$ Gyr using \texttt{stardate}. Repeating the steps we derive an age of 0.9 $\pm$ 0.1 Gyr for WASP-140, using the \texttt{gyro-interp} tool, and 2.9$^{+1.1}_{-2.1}$ Gyr using \texttt{stardate}. Age determinations that rely on different methods (e.g., isochrone fitting, gyrochronology, lithium abundance, and chromospheric activity) can yield variations because each approach is sensitive to a distinct combination of stellar properties and evolutionary stages \citep[e.g.,][]{soder10}. Isochrone fitting is often more robust for stars that have evolved off the Zero-Age Main Sequence. Gyrochronology, on the other hand, leverages the relationship between rotation period and age, which has been calibrated primarily for solar-like dwarfs in clusters of known ages \citep[e.g.,][]{bou24}. As a consequence, this technique can suffer biases for stars that do not follow the standard spin-down tracks (e.g., due to tidal interactions, unusual magnetic activity, or past merging events) and is less reliable for F-type stars, which may spin down more slowly or exhibit weaker magnetic braking \citep{vanSaders16}. Similarly, lithium abundance can provide an age estimate by tracing depletion over time in G/K dwarfs, yet there is significant star-to-star scatter at a given age, and this method is more sensitive in younger populations when Li depletion is rapid \citep[e.g.,][]{jeff23}. Chromospheric activity indices (e.g., $R'_{\rm HK}$) also correlate with stellar age, but the calibration has intrinsic scatter and may be affected by short-term stellar variability and/or episodic activity cycles \citep{mam08}. The differences found for TOI-2046 and Qatar-4 are therefore consistent with the typical systematic limitations of each technique. Nonetheless, the broad agreement among the various indicators is that these systems are older than previously reported.

\section{Future atmospheric characterization}
\label{atmo}

The composition of exoplanetary atmospheres can be used in a complementary way to study the formation and evolution of exoplanetary systems \citep{daw18}. In particular, the elemental ratios (e.g., C/O, N/O, O/H, S/N) can be used to constrain the system's history as well as to investigate the chemically diverse environment in protoplanetary disk \citep[e.g.,][]{tur21,pace22}. Since the launch of the JWST, we have now more than a dozen systems with measured atmospheric composition with an unattainable precision a few years ago. This number will steadily grow in the future. These data will be complemented by the ESA \textit{Ariel} mission \citep{tin18} that will obtain homogeneous spectroscopic data for hundreds of atmospheres \citep{boc23}. We have computed the transmission and emission spectroscopy metric \citep{kemp18} to assess which targets from our sample are the most amenable to future follow-up. As can be seen from Table \ref{tab:met}, the most amenable target for transmission spectroscopy is HAT-P-41 Ab.

\begin{table*}[htbp]
\caption{Transmission (TSM) and emission (ESM) spectroscopy metrics for our sample, as defined in \citet{kemp18}. The higher the value, the higher the suitability for atmospheric characterization.}

\label{tab:met}
\centering
\begin{tabular}{lcccccccc}
\hline
 & W48b & W59b & W140 Ab & W173 Ab & T2046b & HP41 Ab & HP50b & Q4b \\ 
\hline
TSM & 116 & 40 & 130 & 31 & 86 & 416 & 36 & 9 \\
ESM & 91 & 34 & 292 & 139 & 105 & 248 & 48 & 71 \\
\hline
\end{tabular}
\end{table*}

The atmosphere of HAT-P-41 Ab has been studied by \citet{shep21} and \citet{jia24}, who reported one of the highest metallicities of the known hot Jupiters. Such a high metallicity (especially O/H) supports the migration of the planets through the disk via viscous torques over other mechanisms (disk-free migration or formation via pebble accretion whereby the oxygen-rich solids are locked in the core). As discussed in \citet{shep21}, the previously reported moderately misaligned orbit of the HAT-P-41 Ab is in tension with the disk migration hypothesis. HAT-P-41 Ab was initially considered for detailed atmospheric study with JWST, however, due to the reported misalignment it has been withdrawn from the otherwise homogeneous sample \citep{kirk24}. Our value of $\lambda =-4.4^{+5.0}_{-5.6}$\,deg is in better agreement with the disk migration scenario where the planet has formed outside of the $\rm{H_2O}$ snowline and migrated inward while accreting substantial mass in planetesimals. 

\begin{figure}
\includegraphics[width=0.5\textwidth]{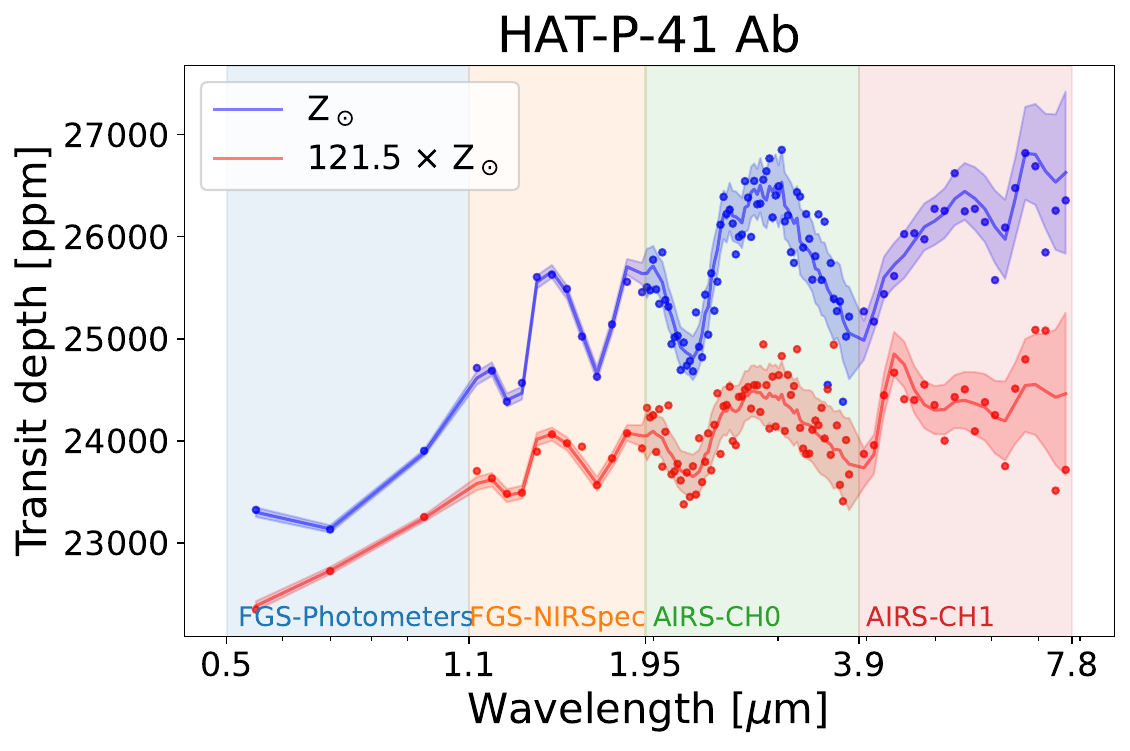}
\caption{Simulated \textit{Ariel} transmission spectrum of HAT-P-41 Ab’s atmosphere. The solid lines represent the unscattered spectra. The shaded colored areas correspond to the 1-$\sigma$ confidence levels of the \textit{Ariel} observations simulated in Tier 3 of the mission \citep{Edwards2019a}. The dots are noisy data representing observed spectra. In red, we show a spectrum with enhanced metallicity, consistent with \citet{shep21}. In blue, we show a spectrum with solar metallicity as derived by \citet{chan22t}. \textit{Ariel} will be able to distinguish between these two models with high confidence with three observed transits due to its broad wavelength coverage and high stability.}
\label{arielhp41}
\end{figure}

However, the enhanced metallicity result was not confirmed by the independent analysis of the HST data by \citet{chan22t} who reported sub-solar to solar metallicity. The main limiting factor when deriving precise metallicity and the C/O ratio with the HST data is the narrow wavelength coverage of the HST and the presence of complex systematics \citep{gib11}. \textit{Ariel} will provide simultaneous coverage from 0.6 to 7.8 $\mu$m, allowing the determination of the chemical composition with high confidence. In Fig. \ref{arielhp41} we provide the \textit{Ariel} simulated transmission spectrum of HAT-P-41 Ab obtained with three transits. We provide \textit{Ariel} spectrum rather than one from JWST as a conservative estimate due to the smaller mirror diameter of \textit{Ariel}. \textit{Ariel} is set to perform a homogeneous study of hundreds of planetary atmospheres. We have used the same setup \citep{mug20,Mugnai2023} as in \citet{zak24a} to simulate the \textit{Ariel} spectra. \textit{Ariel} data will be able to unambiguously distinguish between solar-like and enhanced metallicity. The most prominent features for the metallicity determination are $\rm{H_2O}$ and $\rm{CO_2}$. With a day-side temperature of $\sim$ 2300 K HAT-P-41 Ab lies in the transitional region between hot and ultra-hot Jupiters where physical processes such as molecular dissociation and H$^-$ opacity become important. By understanding the thermal structure of the atmosphere it will be possible to narrow down the onset conditions of these processes. The inferred well-aligned orbit of HAT-P-41 Ab would be consistent with the outcome of disk migration where the formation models expect an oxygen-rich atmosphere with high-atmospheric metallicity \citep{mad14}. High-confidence confirmation of the enhanced metallicity is hence important for testing our current planet formation models. If, on the other hand, sub-solar to solar metallicity is derived, consistently with the results by \citet{chan22t}, this would hint at different processes at play.

\section{Additional figures}
\label{mcmcsec}
\setcounter{figure}{0}
\renewcommand{\thefigure}{E\arabic{figure}}

\begin{figure}[h]
\includegraphics[width=0.5\textwidth]{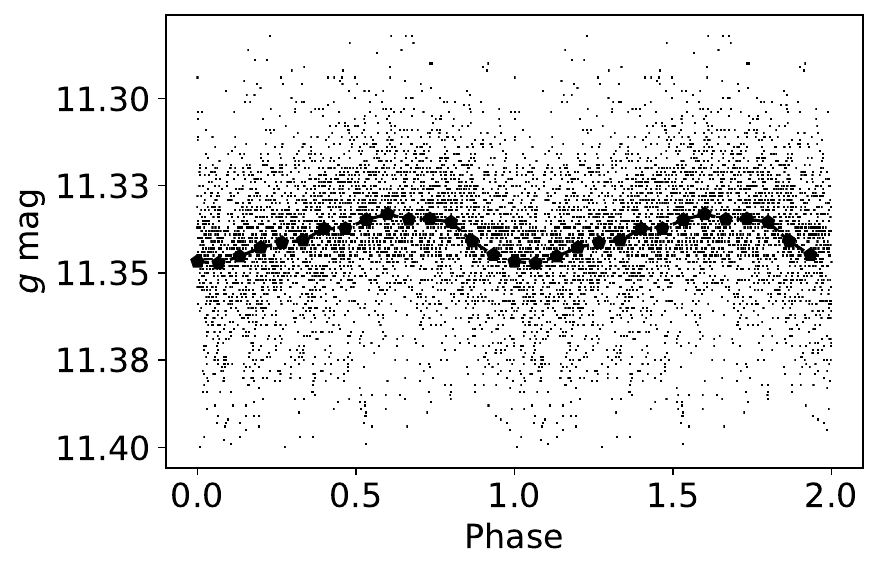}
\caption{Phased $g$-band ASAS-SN data of WASP-140 with the detected stellar rotation period of $P_{\rm{Rot}}$ =10.44 $\pm$ 0.05 d.}
\label{fig:perrot1}
\end{figure}

\begin{figure}[h]
\includegraphics[width=0.5\textwidth]{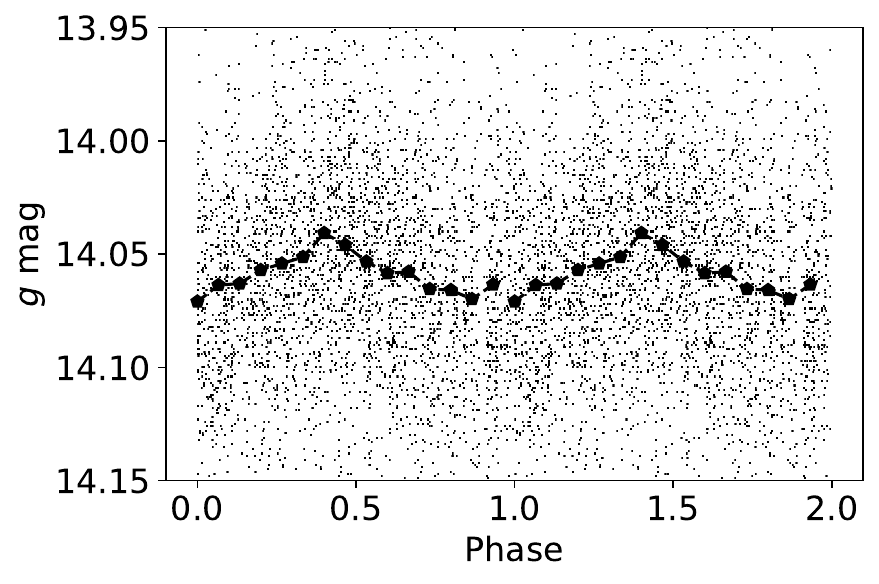}
\caption{Phased $g$-band ASAS-SN data of Qatar-4 with the detected stellar rotation period of $P_{\rm{Rot}}$ =7.07 $\pm$ 0.08 d.}
\label{fig:perrot2}
\end{figure}

\begin{figure}[h]
\includegraphics[width=0.5\textwidth]{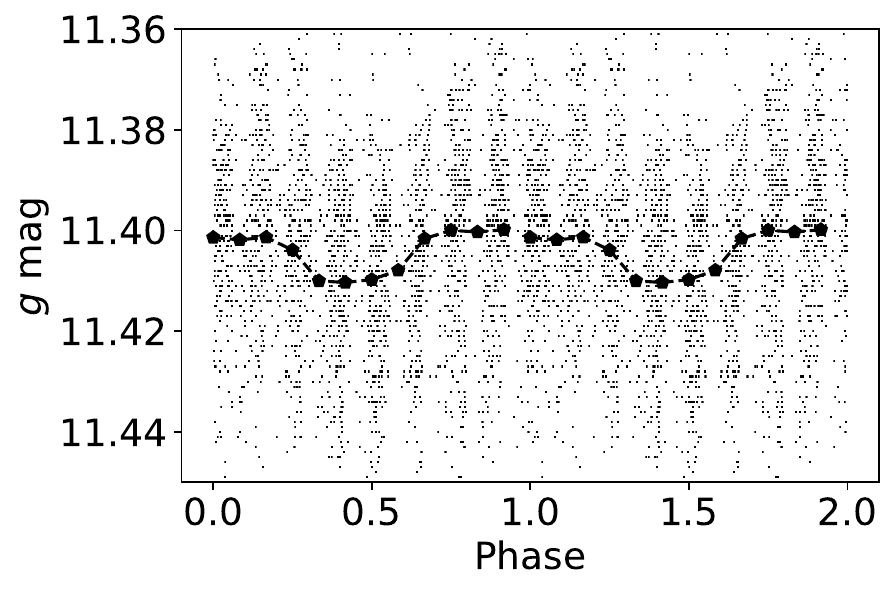}
\caption{Phased $g$-band ASAS-SN data of WASP-173\,A with the detected stellar rotation period of $P_{\rm{Rot}}$ =7.97 $\pm$ 0.05 d.}
\label{fig:perrot3}
\end{figure}

\begin{figure}[h]
\includegraphics[width=0.5\textwidth]{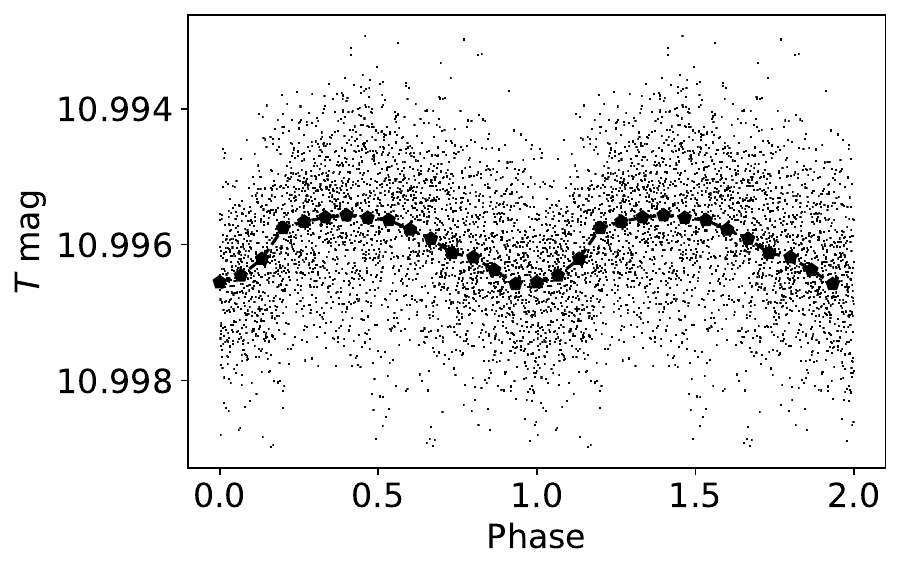}
\caption{Phased \textit{TESS} data of TOI-2046 with the detected stellar rotation period of $P_{\rm{Rot}}$ =4.05 $\pm$ 0.05 d.}
\label{fig:perrot4}
\end{figure}

\begin{figure*}
\includegraphics[width=\textwidth]{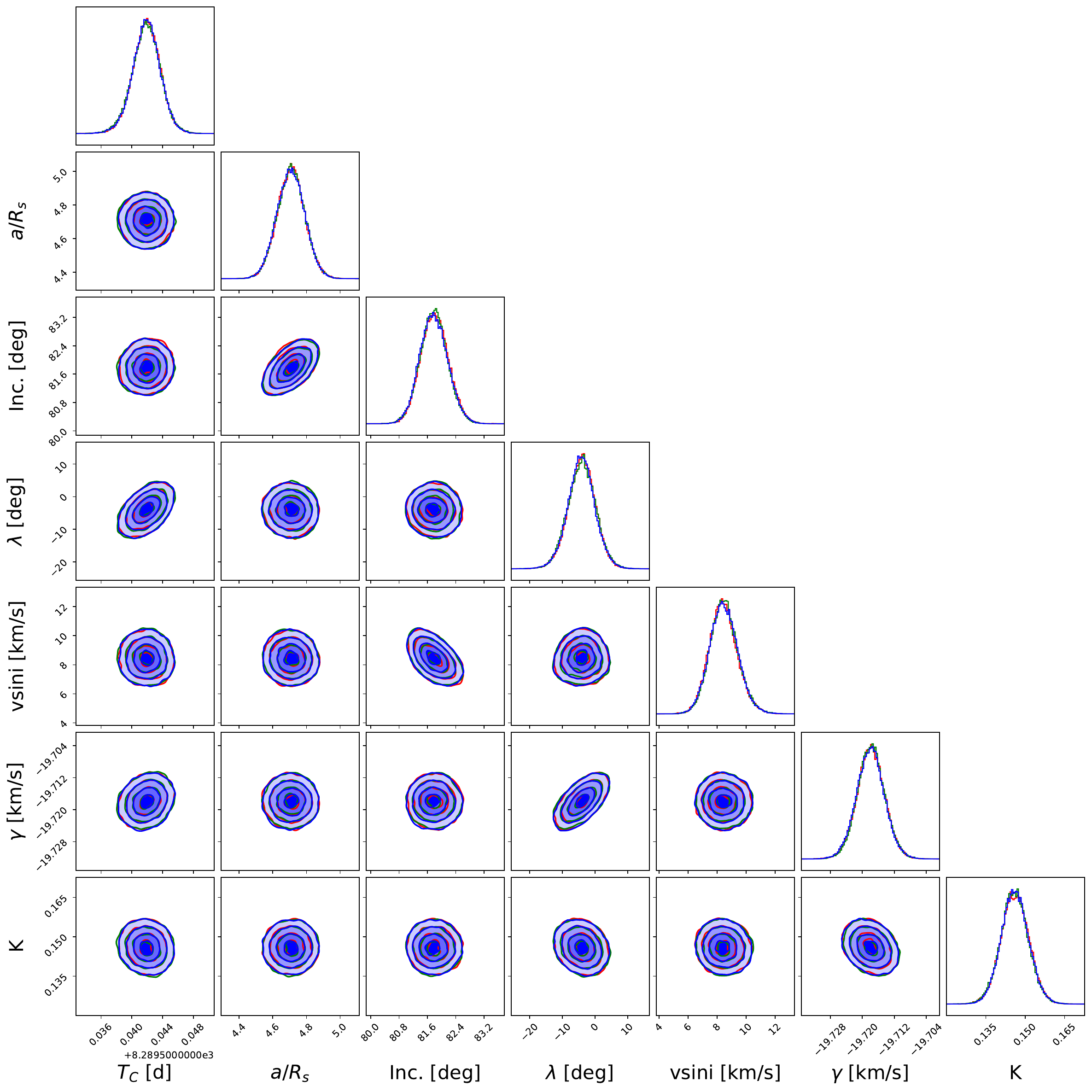}
\caption{MCMC results of WASP-48b. Three independent MCMC simulations are shown with different colors.}
\label{m48}
\end{figure*}

\begin{figure*}
\includegraphics[width=\textwidth]{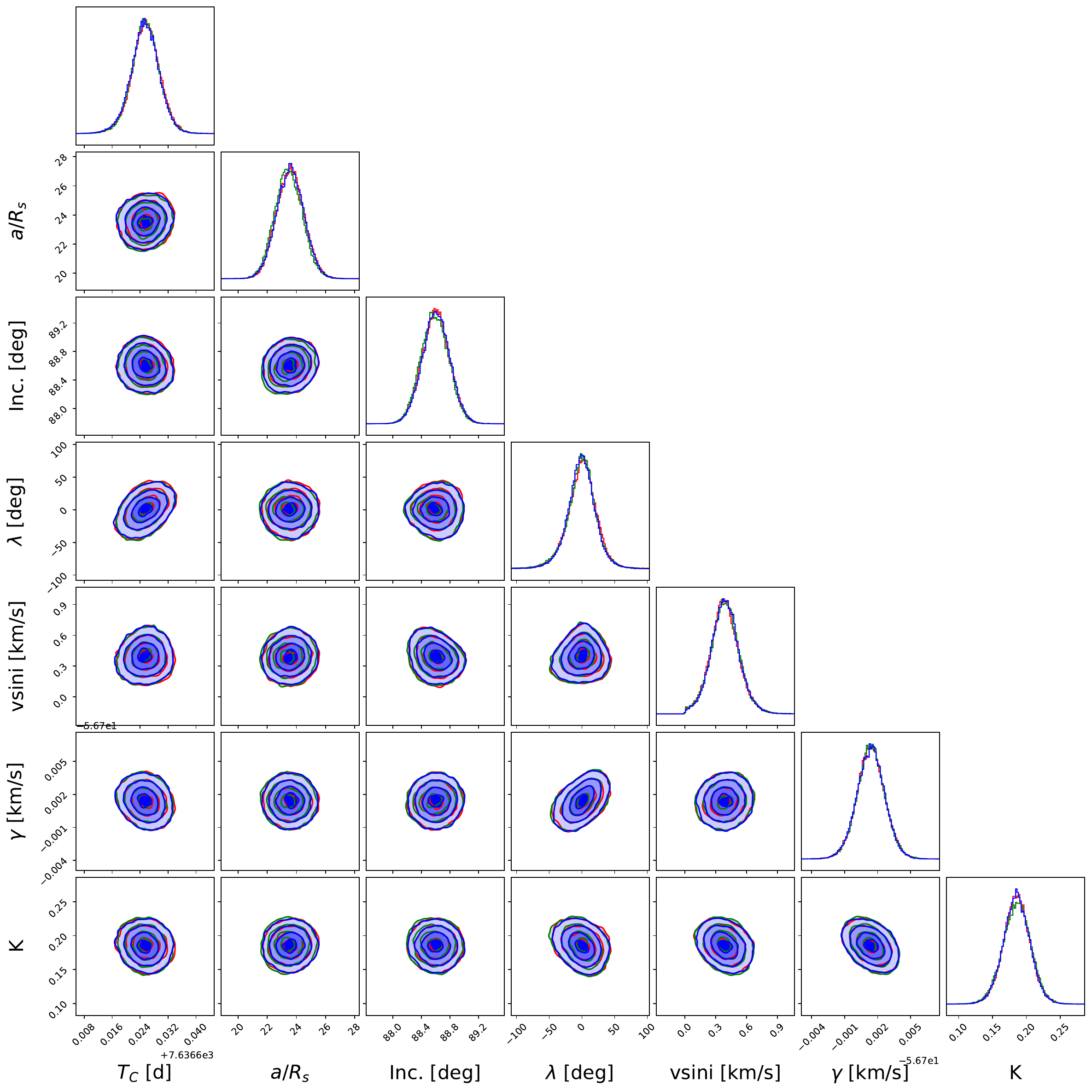}
\caption{MCMC results of WASP-59b. Three independent MCMC simulations are shown with different colors.}
\label{m59}
\end{figure*}

\begin{figure*}
\includegraphics[width=\textwidth]{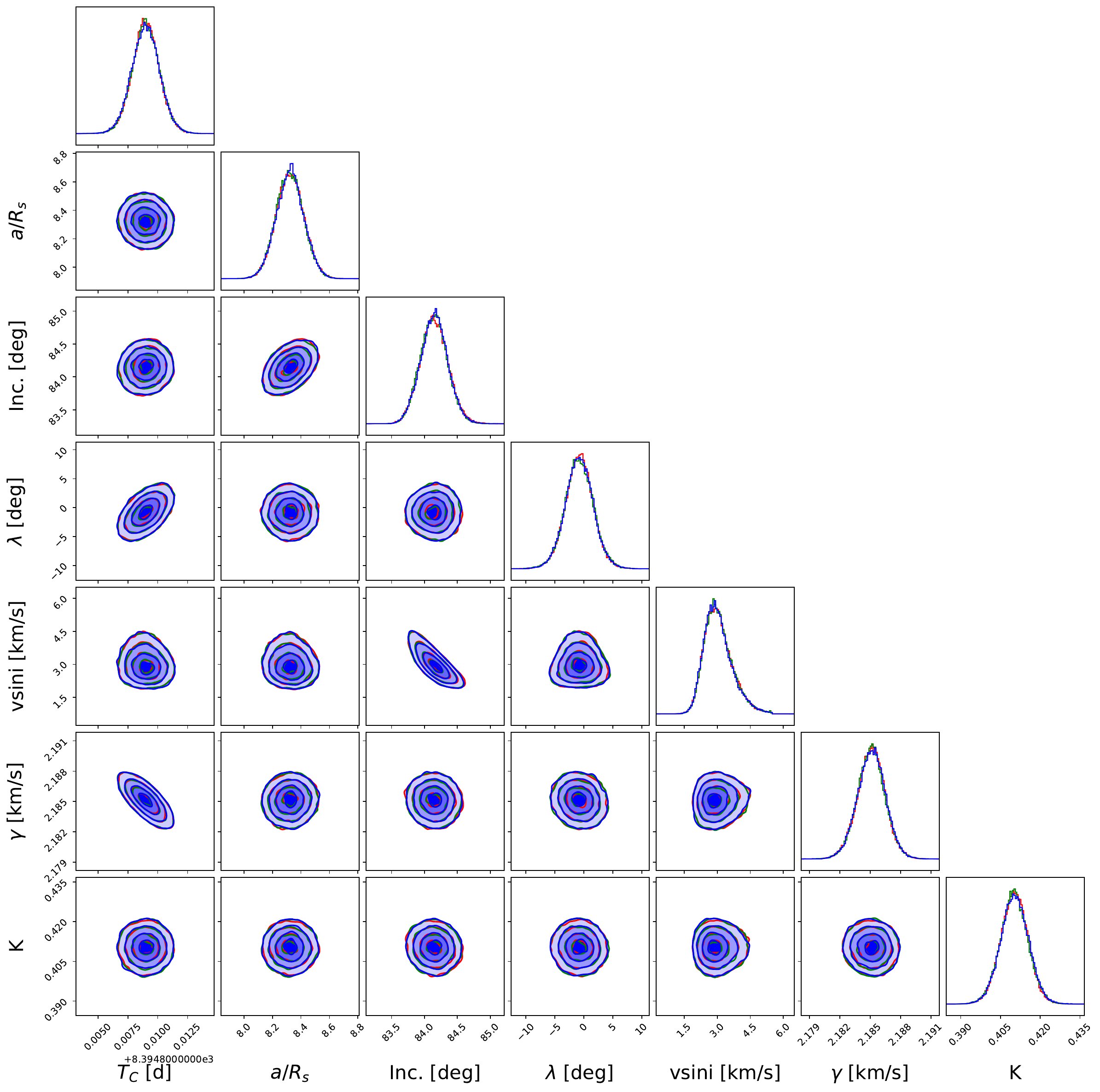}
\caption{MCMC results of WASP-140 Ab. Three independent MCMC simulations are shown with different colors.}
\label{m140}
\end{figure*}

\begin{figure*}
\includegraphics[width=\textwidth]{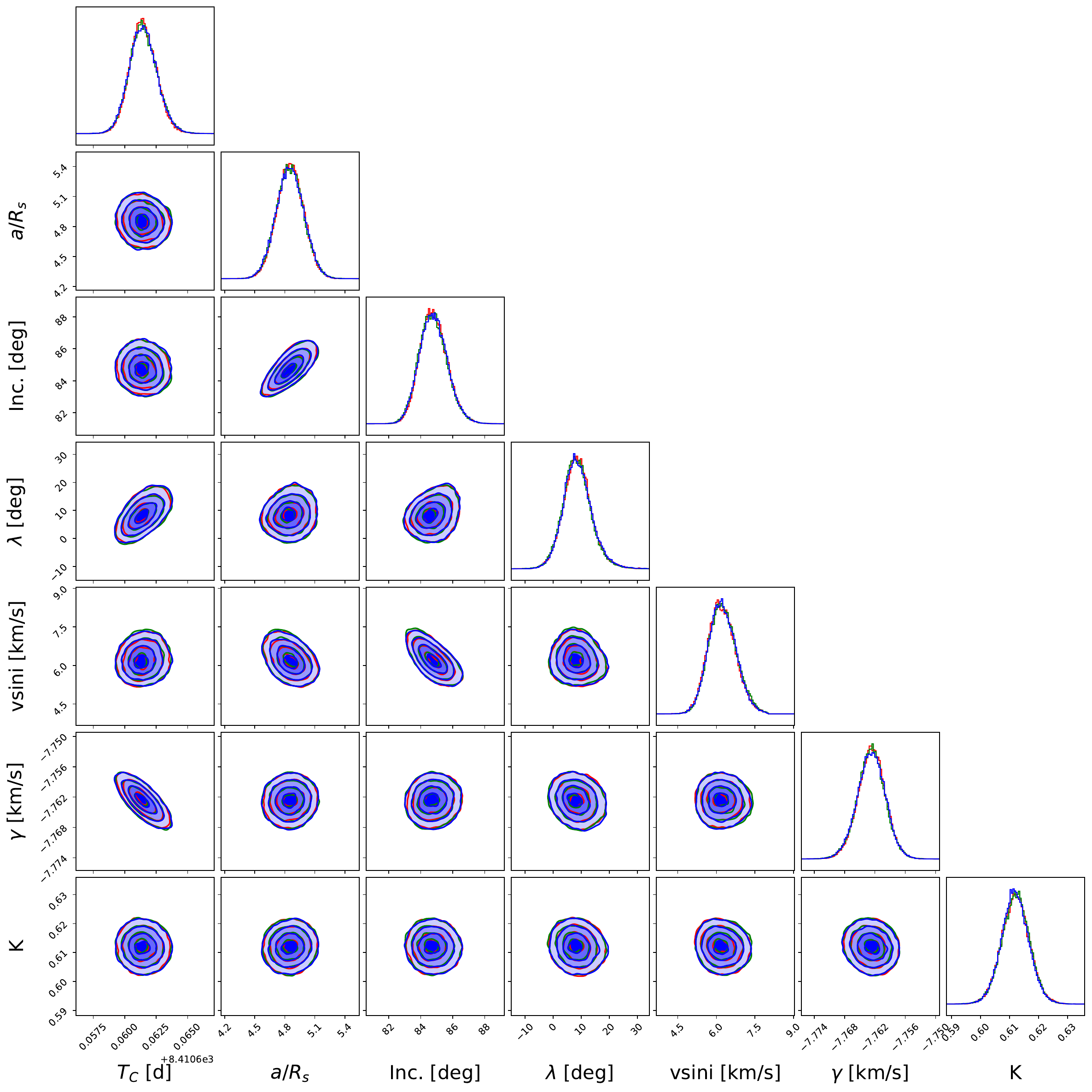}
\caption{MCMC results of WASP-173 Ab. Three independent MCMC simulations are shown with different colors.}
\label{m173}
\end{figure*}

\begin{figure*}
\includegraphics[width=\textwidth]{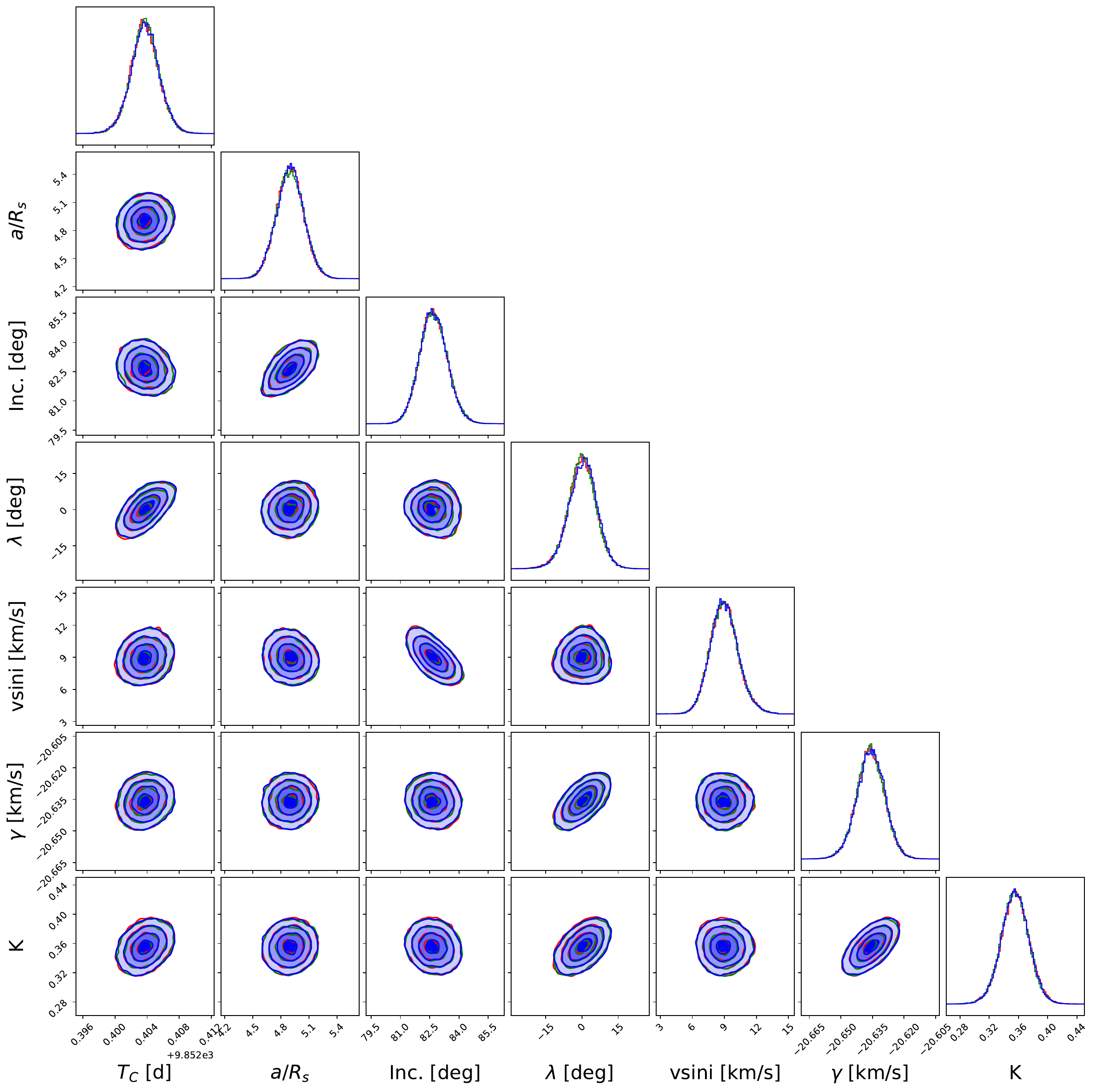}
\caption{MCMC results of TOI-2046b. Three independent MCMC simulations are shown with different colors.}
\label{m2046}
\end{figure*}

\begin{figure*}
\includegraphics[width=\textwidth]{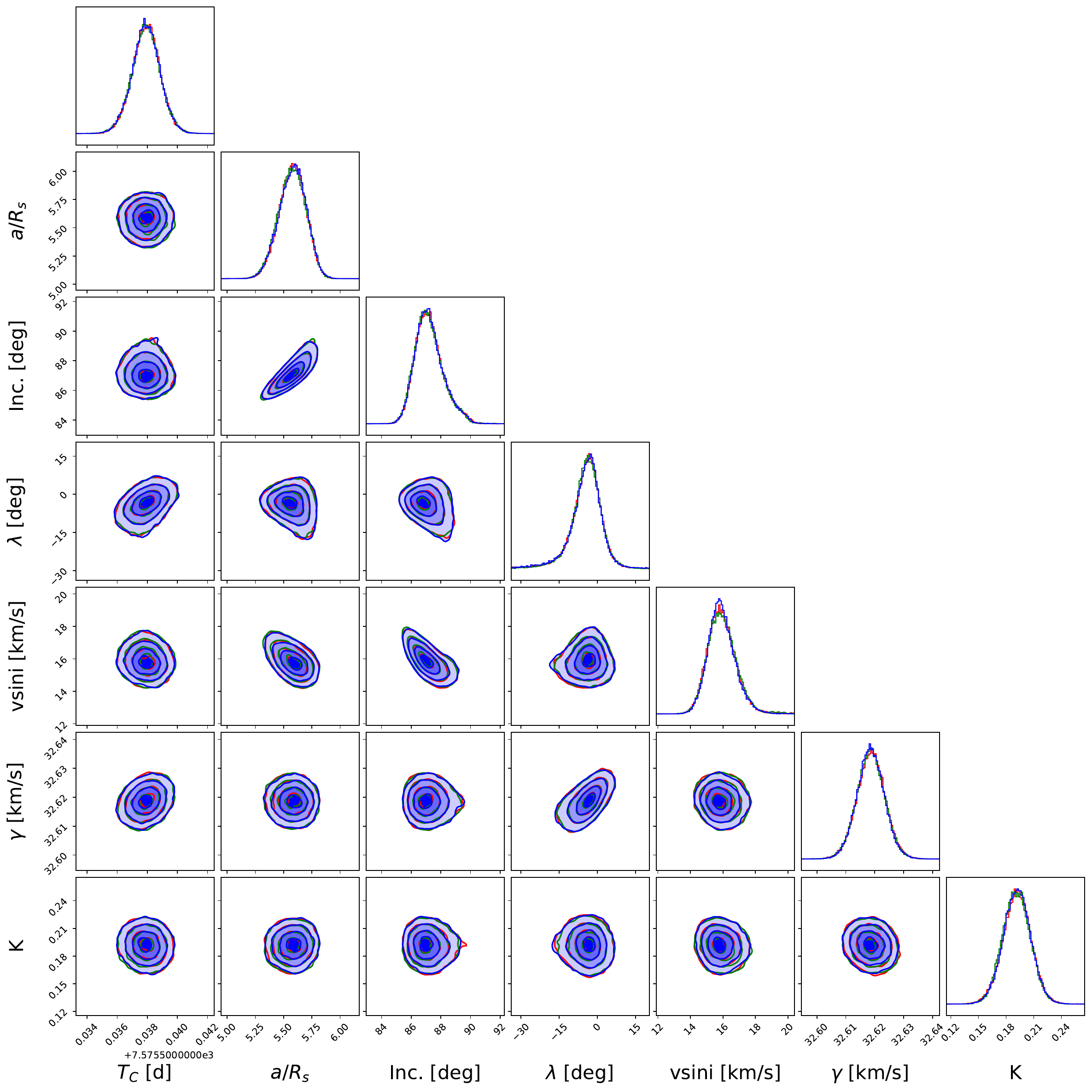}
\caption{MCMC results of HAT-P-41 Ab. Three independent MCMC simulations are shown with different colors.}
\label{mhp41}
\end{figure*}

\begin{figure*}
\includegraphics[width=\textwidth]{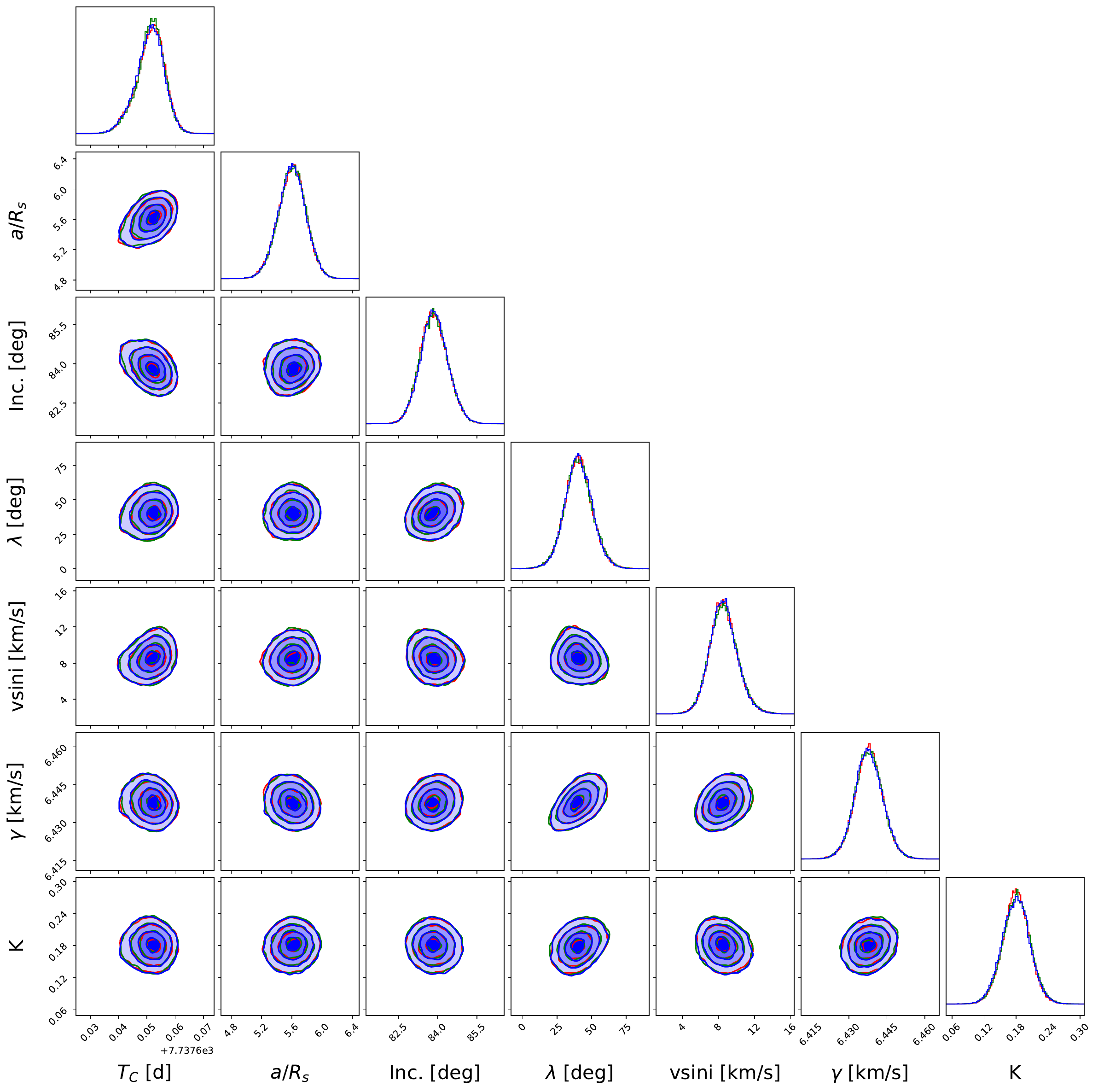}
\caption{MCMC results of HAT-P-50b. Three independent MCMC simulations are shown with different colors.}
\label{mhp50}
\end{figure*}

\begin{figure*}
\includegraphics[width=\textwidth]{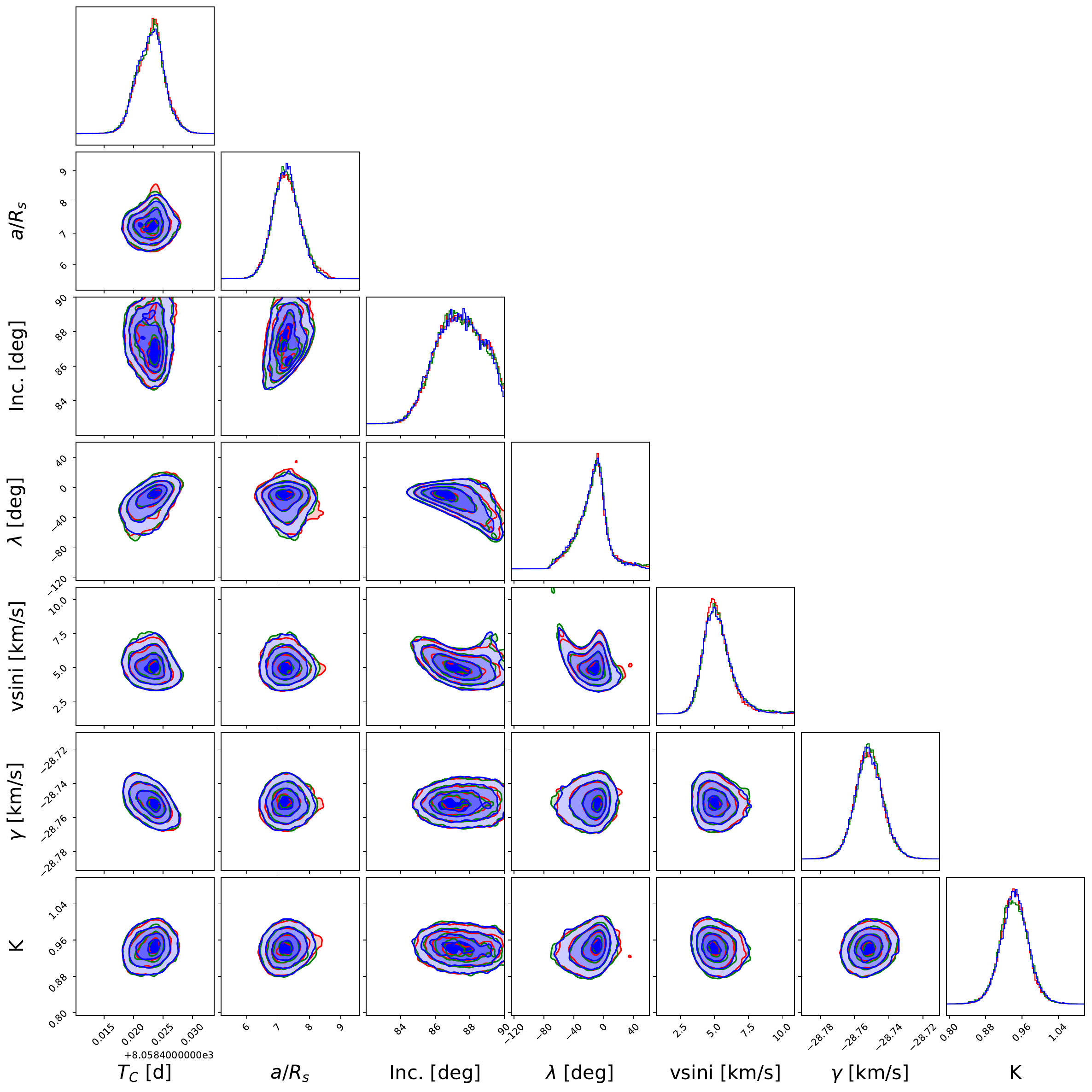}
\caption{MCMC results of Qatar-4b. Three independent MCMC simulations are shown with different colors.}
\label{mq4}
\end{figure*}

\end{document}